\documentclass[sigconf, nonacm]{acmart}
\usepackage{algorithm}
\usepackage{algorithmic}
\usepackage{multirow}
\usepackage{subcaption}
\newcommand\vldbdoi{XX.XX/XXX.XX}
\newcommand\vldbpages{XXX-XXX}
\newcommand\vldbvolume{20}
\newcommand\vldbissue{1}
\newcommand\vldbyear{2026}
\newcommand\vldbauthors{\authors}
\newcommand\vldbtitle{\shorttitle} 
\newcommand\vldbavailabilityurl{https://github.com/oceanbase/oceanbase/tree/4.4.x}
\newcommand\vldbpagestyle{plain} 

\begin{document}
\title{OceanBase Bacchus: a High-Performance and Scalable Cloud-Native Shared Storage Architecture for Multi-Cloud}

\settopmatter{authorsperrow=5}
\makeatletter
\def\@authorfont{\normalsize\sffamily}
\def\@affiliationfont{\normalsize}
\makeatother

\author{Quanqing Xu}
\affiliation{%
  \institution{OceanBase, Ant Group}
}
\authornote{Quanqing Xu and Mingqiang Zhuang contributed equally to this work.}

\author{Mingqiang Zhuang}
\affiliation{%
  \institution{OceanBase, Ant Group}
}
\authornotemark[1]


\author{Chuanhui Yang}
\affiliation{%
  \institution{OceanBase, Ant Group}
}
\authornote{Chuanhui Yang is the corresponding author.}

\author{Quanwei Wan}
\affiliation{%
  \institution{OceanBase, Ant Group}
}

\author{Fusheng Han}
\affiliation{%
  \institution{OceanBase, Ant Group}
}

\author{Fanyu Kong}
\affiliation{%
  \institution{OceanBase, Ant Group}
}

\author{Hao Liu}
\affiliation{%
  \institution{OceanBase, Ant Group}
}

\author{Hu Xu}
\affiliation{%
  \institution{OceanBase, Ant Group}
}

\author{Junyu Ye}
\affiliation{%
  \institution{OceanBase, Ant Group}
  \vspace{0.5em}
  \hspace{-33em}\institution{OceanBaseLabs@service.oceanbase.com}
}






\begin{abstract}
Although an increasing number of databases now embrace shared-storage architectures, current storage-disaggregated systems have yet to strike an optimal balance between cost and performance. In high-concurrency read/write scenarios, B+-tree--based shared storage struggles to efficiently absorb frequent in-place updates. Existing LSM-tree--backed disaggregated storage designs are hindered by the intricate implementation of cross-node shared-log mechanisms, where no satisfactory solution yet exists. 

This paper presents OceanBase Bacchus, an LSM-tree architecture tailored for object storage provided by cloud vendors. The system sustains high-performance reads and writes while rendering compute nodes stateless through shared service-oriented PALF (Paxos-backed Append-only Log File system) logging and asynchronous background services. We employ a Shared Block Cache Service to flexibly utilize cache resources. Our design places log synchronization into a shared service, providing a novel solution for log sharing in storage-compute-separated databases. The architecture decouples functionality across modules, enabling elastic scaling where compute, cache, and storage resources can be resized rapidly and independently. Through experimental evaluation using multiple benchmark tests, including SysBench and TPC-H, we confirm that OceanBase Bacchus achieves performance comparable to or superior to that of HBase in OLTP scenarios and significantly outperforms StarRocks in OLAP workloads. Leveraging Bacchus’s support for multi-cloud deployment and consistent performance, we not only retain high availability and competitive performance but also achieve substantial reductions in storage costs—by 59\% in OLTP scenarios and 89\% in OLAP scenarios.
\end{abstract}

\maketitle

\pagestyle{\vldbpagestyle}
\begingroup\small\noindent\raggedright\textbf{PVLDB Reference Format:}\\
\vldbauthors. \vldbtitle. PVLDB, \vldbvolume(\vldbissue): \vldbpages, \vldbyear.\\
\href{https://doi.org/\vldbdoi}{doi:\vldbdoi}
\endgroup
\begingroup
\renewcommand\thefootnote{}\footnote{\noindent
This work is licensed under the Creative Commons BY-NC-ND 4.0 International License. Visit \url{https://creativecommons.org/licenses/by-nc-nd/4.0/} to view a copy of this license. For any use beyond those covered by this license, obtain permission by emailing \href{mailto:info@vldb.org}{info@vldb.org}. Copyright is held by the owner/author(s). Publication rights licensed to the VLDB Endowment. \\
\raggedright Proceedings of the VLDB Endowment, Vol. \vldbvolume, No. \vldbissue\ %
ISSN 2150-8097. \\
\href{https://doi.org/\vldbdoi}{doi:\vldbdoi} \\
}\addtocounter{footnote}{-1}\endgroup

\ifdefempty{\vldbavailabilityurl}{}{
\vspace{.3cm}
\begingroup\small\noindent\raggedright\textbf{PVLDB Artifact Availability:}\\
The source code, data, and/or other artifacts have been made available at \url{\vldbavailabilityurl}.
\endgroup
}

\section{Introduction}
The database industry is rapidly embracing cloud-native architectures, with ``Public Cloud First'' becoming the prevailing trend. Shared storage architectures enable cloud-native databases by eliminating data redundancy~\cite{10.1145/3035918.3056101}, simplifying consistency management~\cite{10.1145/3299869.3314047}, and enabling elastic scaling~\cite{10.1145/2882903.2903741}. However, existing shared storage solutions face fundamental challenges for high-concurrency online transaction processing (OLTP) workloads.

\textbf{The Challenge of B+-tree--based Shared Storage.} B+-tree--based shared storage struggles with high-concurrency writes. Random I/O patterns and in-place updates create write conflicts when multiple compute nodes access shared storage. Frequent page splits and tree rebalancing degrade performance under write-intensive workloads, making B+-tree architectures ill-suited for cloud-native applications requiring high write throughput.

\textbf{The Challenge of LSM-tree--based Shared Storage.} While LSM-trees offer superior write performance, existing LSM-tree--backed disaggregated storage faces a critical bottleneck: the lack of efficient cross-node shared-log mechanisms. Systems persisting logs directly in durable storage suffer from high latency, while those relying on log replication struggle with consistency and scalability. This challenge has hindered LSM-tree adoption in shared storage, despite their alignment with object storage's append-only nature.

\textbf{The Need for Object Storage Integration.} Cloud vendors provide low-cost object storage (e.g., AWS S3, Alibaba Cloud OSS (Object Storage Service)) that is 85\% cheaper than traditional cloud disks. However, object storage's high latency and limited IOPS (I/O operations per second) challenge OLTP workloads requiring millisecond response times. Existing systems either avoid object storage or use it only for cold data, missing substantial cost savings.

\begin{figure}[htp]
    \centering
    \includegraphics[trim=0 0 0 0,clip,width=0.99\linewidth]{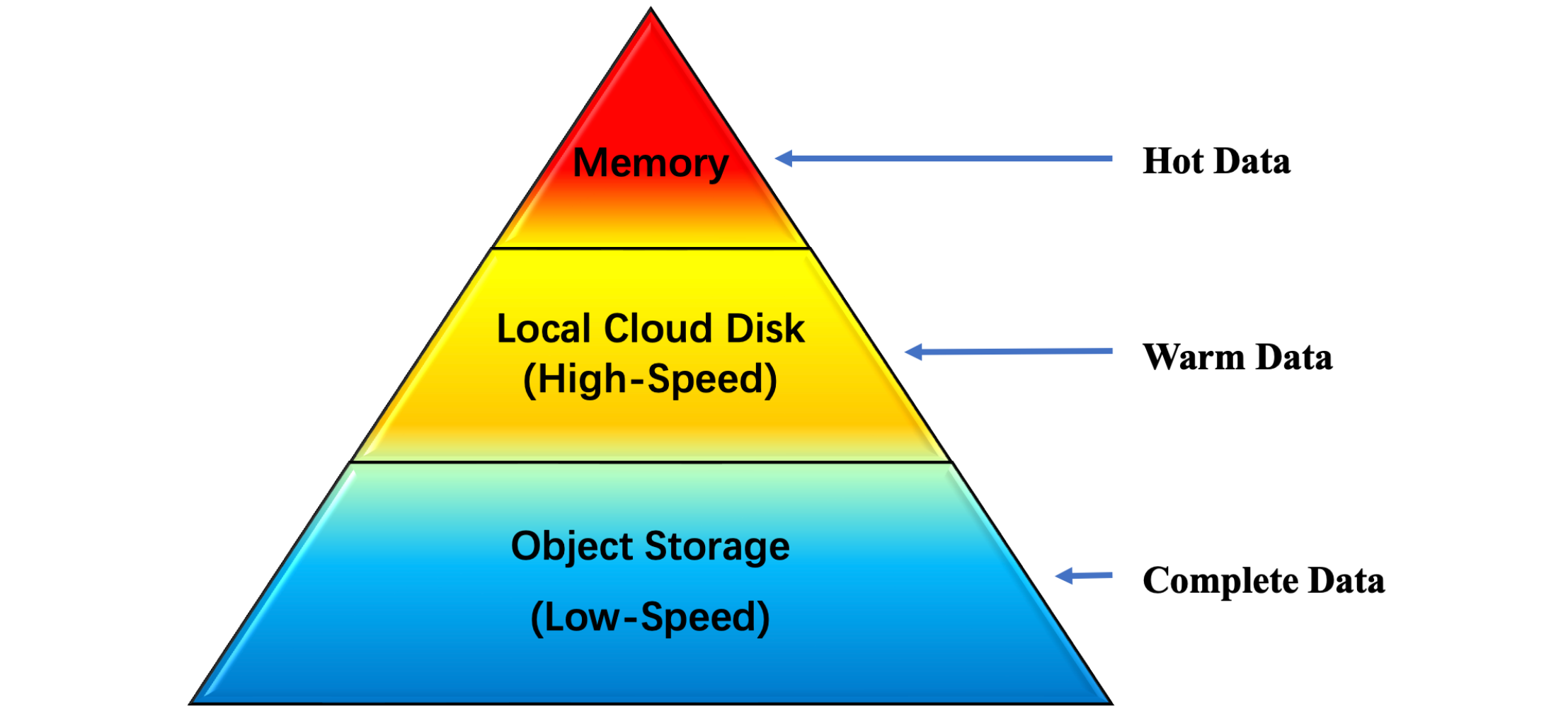}
    \caption{An illustration of hot/cold data storage}
    \label{fig:OBHotData}
    \vspace*{-6pt}
\end{figure}

\textbf{The Motivation for OceanBase Bacchus.} We need a shared storage architecture that: (1) leverages LSM-tree's write-optimized design while solving shared-log coordination, (2) integrates low-cost object storage without sacrificing OLTP performance, (3) enables storage-compute disaggregation with stateless compute nodes, and (4) supports flexible multi-cloud deployment. This paper focuses on \textbf{OLTP} as the primary target, with distinct hot/cold data patterns (historical databases, time-series, multi-model), where shared storage benefits are most pronounced. The same architecture also supports \textbf{online analytical processing (OLAP)} and \textbf{HTAP} (hybrid OLTP and OLAP) workloads; we evaluate both OLTP and OLAP in the experiments. Figure~\ref{fig:OBHotData} illustrates hot/cold data storage.

\textbf{Our Solution: OceanBase Bacchus.} To address these challenges, we propose OceanBase Bacchus\footnote{Bacchus was considered a deity associated with the sharing spirit in ancient Roman mythology, so it is used to name our shared storage database, i.e., OceanBase 4.4.x.}, an LSM-tree~\cite{10.1007/s002360050048} architecture specifically designed for object storage provided by cloud vendors. Our design addresses the fundamental limitations of existing shared storage systems through three key innovations.

First, we solve shared-log coordination through a service-oriented PALF-backed logging architecture~\cite{10.14778/3685800.3685803}. Logging is restructured into an independent service with PALF consensus, enabling efficient cross-cluster log sharing. Both data logs and metadata logs (\textit{journal}) are stored in this service. Compared to shared-nothing architectures~\cite{10.1145/3318464.3386134}, this eliminates log redundancy. Compared to persisting logs directly in durable storage~\cite{10.1145/3318158}, our log service eliminates real-time I/O on shared storage, reducing latency.

Second, we bridge the performance gap through multi-layer caching (local persistent cache + distributed cache). Hot data is cached in low-latency local storage, while cold data leverages object storage's cost advantages. Our custom LSM-tree engine adapts to object storage's append-only nature, avoiding in-place updates and write conflicts. We optimize I/O aggregation, asynchronous writes, and concurrency control to improve throughput.

Third, we enable cloud-native elasticity through shared services. The system supports serverless architecture and stateless compute nodes via shared cache and asynchronous background services (compaction, DDL, backup, recovery). Persistent cache scaling adapts to hot data fluctuations, reducing costs while ensuring high performance. The architecture supports multi-cloud deployment (AWS, Alibaba Cloud, etc.), avoiding vendor lock-in. 


In summary, OceanBase Bacchus addresses shared storage challenges through: (1) a service-oriented shared-log architecture solving cross-node log coordination, (2) multi-layer caching bridging object storage and OLTP performance, and (3) cloud-native elasticity enabling storage-compute disaggregation. The system provides stateless compute nodes, reduced storage costs (59\% in OLTP, 89\% in OLAP), and high performance. To our knowledge, we are the first to integrate these features into a unified high-performance, low-cost system.

This paper is organized as follows. \textsection \ref{sec:arch} describes the new design of the Bacchus architecture. \textsection \ref{sec:impl} provides an intuitive overview of the implementation of Bacchus shared storage. \textsection \ref{sec:OBIncrement} presents the increment and compaction strategies. \textsection \ref{sec:caching} describes the multi-layer caching and preheating mechanisms. \textsection \ref{sec:GC} details garbage collection mechanisms. We present the experiments in \textsection \ref{sec:eval} to prove the high availability and competitive performance of the ``LSM-tree + Object Storage'' framework. We discuss lessons learned in \textsection \ref{sec:disc}, and review the related work in \textsection \ref{sec:work}. Finally, we state the conclusions in \textsection \ref{sec:conc}. Additionally, OceanBase Bacchus is an open-source project under Mulan Public License 2.0~\cite{mulan} and the source code referenced in this paper is available on both Gitee~\cite{obgitee} and GitHub~\cite{obgithub}.

\section{System Architecture} \label{sec:arch}

\subsection{Layered Design}

\begin{figure*}[htp]
  \centering
  \includegraphics[trim=0 0 0 0,clip,width=0.99\linewidth]{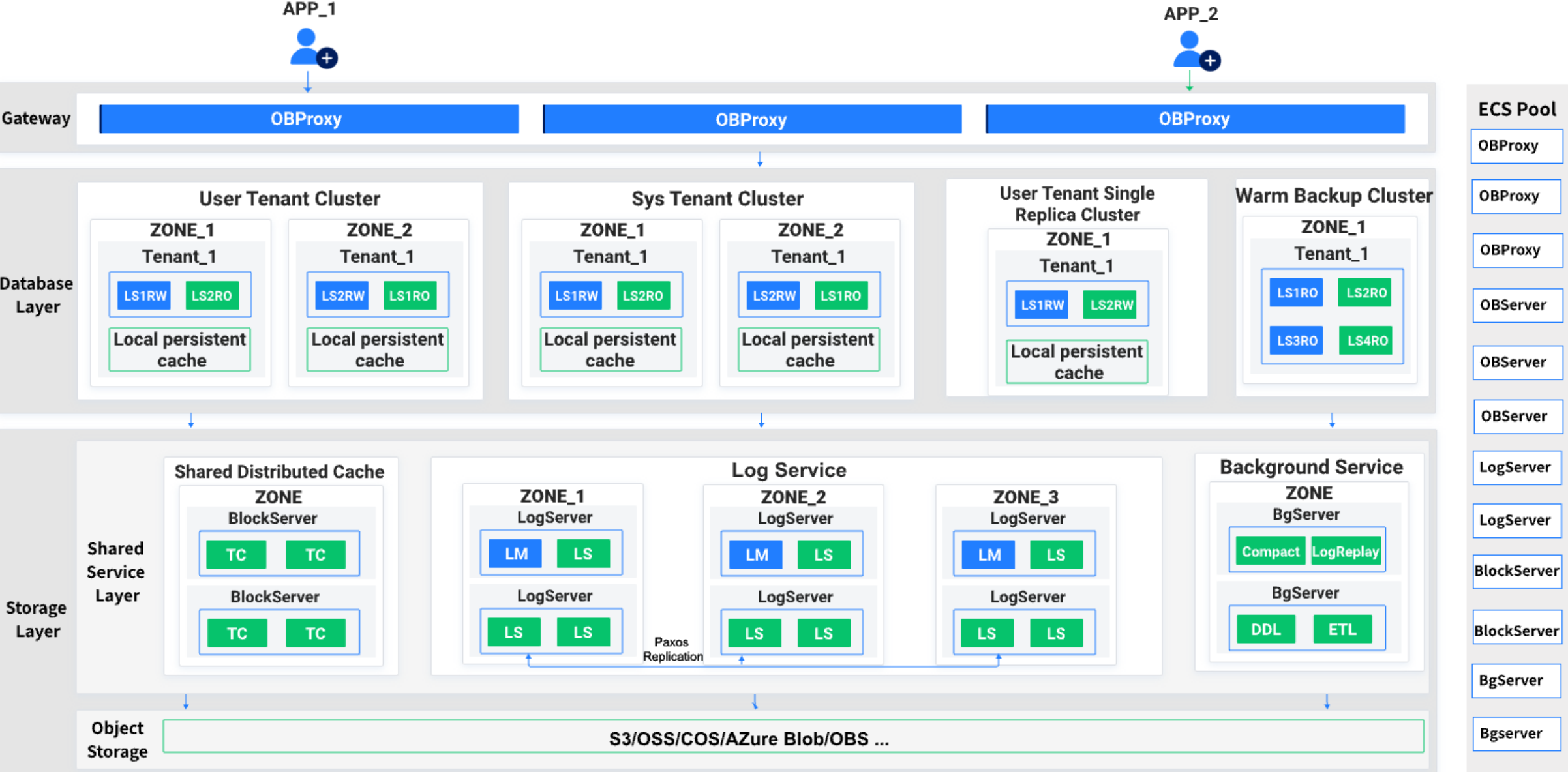}
  \caption{System architecture}
  \label{fig:SystemArchitecture}
  \vspace{-6pt}
\end{figure*}

As shown in Figure~\ref{fig:SystemArchitecture}, the entire architecture is divided into two layers: 1) the Database Layer and 2) the Shared Storage Layer. The Database Layer includes Memory Cache and Local Persistent Cache (more details in \textsection \ref{sec:caching}), while the Shared Storage Layer consists of the Shared Service Layer and Object Storage.


\noindent$\bullet\ $\textbf{Memory Cache} stores the hottest data, including micro-blocks and index data, to ensure that key transactions can respond within milliseconds.

\noindent$\bullet\ $\textbf{Local Persistent Cache} improves access efficiency and serves as the first guarantee of data persistence. It stores hot data, dumped data, warm-up data, and relevant metadata, addressing performance bottlenecks when object storage performs random read operations. Its main functions include:

\textbf{(1) Data Caching:} Caching data read from object storage.

\textbf{(2) Data Prefetching:} Preemptively caching data that is about to be read during operations such as full table scans to enhance data access performance.

\textbf{(3) Data Warming:} Loading hot data into local cache before version merges. The cache remains effective after brief downtime, reducing performance fluctuations. The access sequence is synchronized to replicas in real-time to maintain consistency. During replication, local cache data is migrated to ensure performance.

\textbf{(4) Write Cache:} Utilizing the low-latency characteristics of local storage devices, this layer records essential local state information, e.g., metadata files, the latest change logs, and mini dumps.

\noindent$\bullet\ $\textbf{Shared Service Layer} provides shared services that enable cross-cluster resource sharing and coordination. It consists of three key components:

\textbf{(1) Distributed Cache} is provided by the Shared Block Cache Service, which is implemented by BlockServer nodes that store and serve macro-blocks. It adds critical support to the shared storage architecture by constructing a three-level caching system for hot data, enabling cache warming functionality.

\textbf{(2) Shared Log (Service-Oriented Logging)} is a log service hosted by LogServer nodes, with three independently deployed replicas supporting parallel operation across clusters. It stores metadata and data logs, with replicas achieving consensus via PALF.

\textbf{(3) Background Tasks} such as baseline compaction, DDL, and garbage collection are executed transparently without impacting foreground performance.

\noindent$\bullet\ $\textbf{Object Storage} stores large volumes of infrequently accessed data. Compatible with AWS S3~\cite{s3} interface, supporting multi-cloud solutions (AWS S3, Alibaba Cloud OSS~\cite{alioss}, Blob~\cite{blob}) and local deployments (MinIO~\cite{minio}).

The architecture enables efficient and reliable data retrieval. Unlike PolarDB~\cite{10.14778/3611540.3611562} and GaussDB~\cite{gaussdb}, which are tightly coupled to specific cloud platforms, OceanBase Bacchus supports flexible multi-cloud deployment, avoiding vendor lock-in.

\subsection{Interaction between RW and RO Nodes} \label{sec:rwro}
OceanBase Bacchus consists of the database layer and the shared storage layer. The database layer writes data and metadata to shared storage, where data is stored and metadata is updated accordingly. The system supports multiple RW nodes operating concurrently: data is partitioned into log streams, with each log stream having a single leader responsible for writing to the PALF log service. This sharded write model ensures no data overlap between concurrent writers, eliminating write conflicts while enabling horizontal scaling. Cross-partition transactions are coordinated through distributed 2PC (two-phase commit). The shared process between Read-Write (RW) and Read-Only (RO) nodes is:

\textbf{(1)} RW writes the WAL (Write-Ahead Logging) to the log service during transaction execution.

\textbf{(2)} RO continuously/periodically pulls the WAL from the log service and replays it into local storage.

\textbf{(3)} RW continuously/periodically writes data to shared storage based on the requirements for sharing.

\textbf{(4)} RW continuously/periodically writes metadata updates as journals to the log service based on the requirements for sharing.

\textbf{(5)} RO continuously/periodically pulls journals from the log service and replays them in local storage, replacing metadata in local storage.

\textbf{(6)} RO pulls the newly generated SSTable (Sorted String Tables)~\cite{10.1145/1365815.1365816} list from shared storage based on the updated metadata.

\textbf{(7)} RW continuously/periodically generates metadata based on the journal to bring up new nodes for potential recovery or node scaling.

\subsection{Architecture Features}

This architecture adopts a storage-compute disaggregation design to achieve independent expansion and contraction of computing and storage resources, enabling flexible adaptation to changes in business load. At the data storage level, new CLog (Commit Log) entries are first recorded in the local cloud disk to ensure low-latency update operations, while historical CLog files are migrated to shared storage according to policy, balancing performance and cost efficiency. All data is stored uniformly in shared storage, with hot data cached in local cloud disk for quick access. The cache strategy is dynamically optimized through an automatic identification mechanism, reducing storage redundancy while ensuring high concurrent read/write performance.

The system further incorporates a serverless architecture design that offloads background tasks such as compaction and DDL to independent processes. This approach not only reduces interference with foreground business but also improves computing resource utilization. For performance assurance, the system employs comprehensive preheating mechanisms, smooth major version switching, replication migration, and disaster recovery of master and backup replicas. A Warm Backup Cluster can be maintained in a separate zone to provide a warm standby that continuously replays logs for fast failover. These mechanisms ensure that core business remains stable during architecture upgrades or fault switching, minimize impact on users, and achieve a balance between high availability and high performance.

\subsection{Architecture Advantages}
Due to the innovative, self-developed ``LSM-tree + Object Storage'' architecture and its specialized optimizations, OceanBase Bacchus possesses distinct advantages over other distributed databases. 

\noindent$\bullet\ $\textbf{Coordination between LSM-tree and Object Storage}. Compared to B+-tree--based~\cite{10.1145/356770.356776} systems, B+ trees excel in read performance ($O(logN)$) but suffer from random I/O, page splits, and storage fragmentation under high-concurrency writes. LSM-trees employ tiered storage (MemTable + multi-level SSTables), offering superior write performance through sequential writes and tiered compaction. This aligns with object storage's append-only nature, significantly reducing storage costs.

\noindent$\bullet\ $\textbf{Separation of Functional Modules}. We leverage LSM-tree's tiered nature to offload long-running tasks (e.g., major compaction) to underutilized resources, realizing storage-compute disaggregation. This decouples foreground and background operations. The shared distributed caching scheme isolates caching from compute nodes, enabling independent elastic scaling. A serverless design offloads background tasks to dedicated processes, eliminating over-provisioning and reducing foreground interference.

\noindent$\bullet\ $\textbf{Optimization of Dumping Data}. Unlike Socrates~\cite{10.1145/3299869.3314047} (B-Tree + Shared-disk) and SingleStore~\cite{10.1145/3514221.3526055} (LSM-tree + Blob Store with high synchronization latency), Bacchus's ``LSM-tree + Object Storage'' architecture achieves RPO=0 through stateless compute nodes and fast incremental persistence. RPO (Recovery Point Objective) denotes the maximum acceptable duration of data loss (in time), representing the point to which data must be recovered after a failure. An RPO of 0 means no data loss is acceptable, requiring real-time or near-real-time data replication. We optimize dumping via minute-level and continuous dumping strategies, and introduce \textit{micro compaction} (\textsection~\ref{sec:OBIncrement}) for complex OLTP workloads.

\noindent$\bullet\ $\textbf{Reduction of Storage Costs}.
This architecture achieves significant optimization of storage resources by reducing data copies from 3 to 1 and leveraging the elastic capacity characteristics of object storage. Compared with traditional cloud disks (PL1), OSS costs only 15\% per unit, requires no pre-purchased capacity, and can be expanded on demand. The cost saving effect can be quantified by the following formula:
\begin{equation}
Save = \frac{1 \times N}{( 0.15 + P \times 1 \times N ) \times S }
\label{equ1}
\end{equation}
where $S$ denotes spatial utilization, $P$ represents the hot data ratio, and $N$ is the replica count. Taking spatial utilization $S=80\%$ and replica count $N=3$ as an example:
\begin{enumerate}
\item When the hot data ratio $P=10\%$, the storage cost is reduced by a factor of 8.33.
\item If $P=20\%$, the reduction ratio remains as high as 5$\times$, significantly outperforming traditional solutions.
\item As the hot data ratio increases to $P=50\%$, the reduction ratio decreases to 2.27$\times$ yet still far exceeds conventional approaches.
\end{enumerate}

This design dynamically adapts to business workloads, balancing cost efficiency with data reliability while reducing redundant storage overhead. Additionally, to enhance resource utilization, the cloud disk space allocation for commit logs (CLog) has been compressed from the original 3$\times$ memory size to 1:1 equivalence with memory, further freeing up storage resources. Combined with the low-cost nature of OSS, these optimizations enable the system to guarantee low-latency access to hot data in high-concurrency scenarios while achieving an optimal balance between storage cost and performance through flexible elastic storage capabilities.

\section{System Implementation} \label{sec:impl}

\subsection{Overview}
\begin{figure}[htp]
    \centering
    \includegraphics[width=1.0\linewidth]{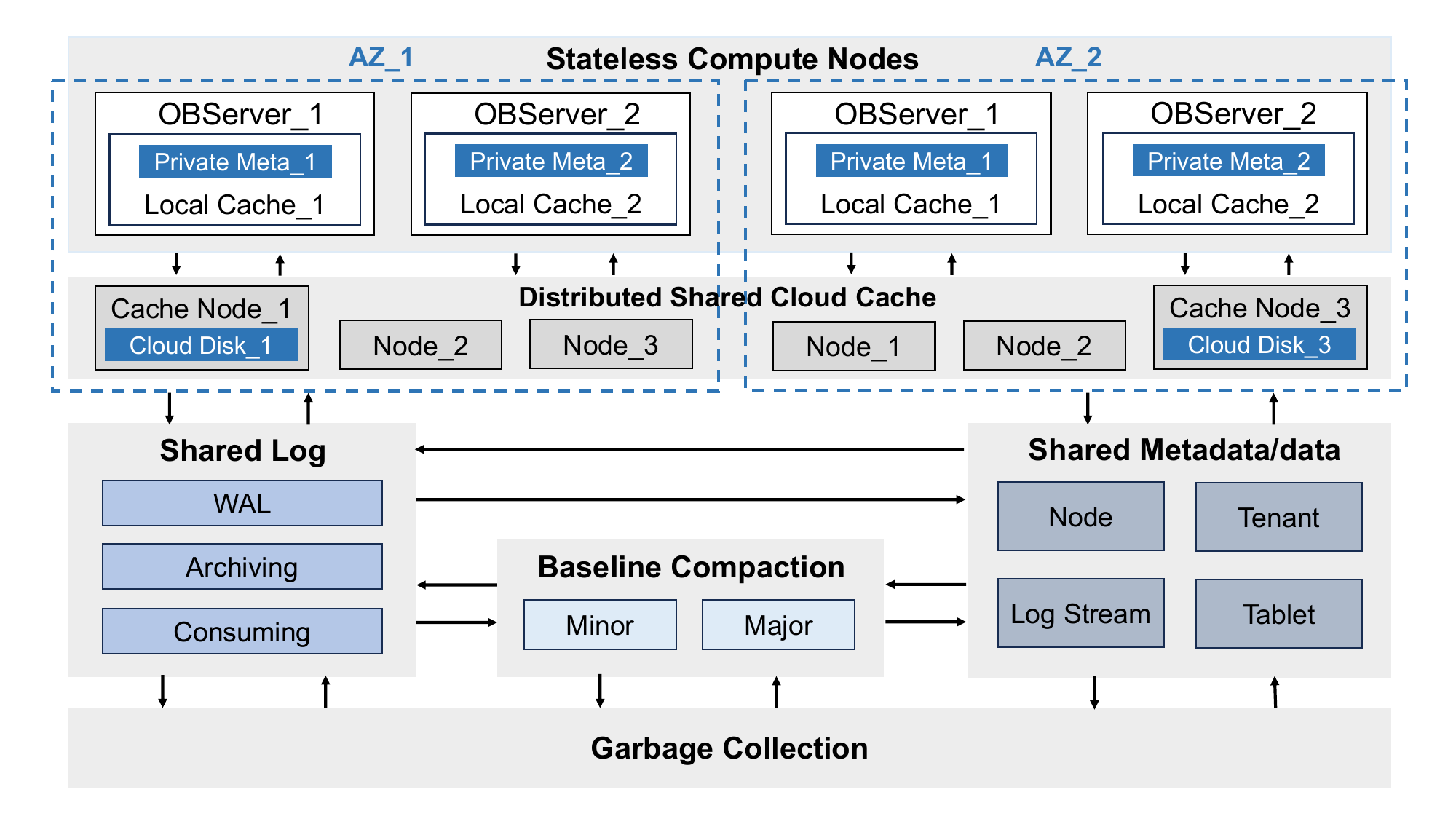}
    \caption{The information interaction among main modules}
    \label{fig:OBModules}
    \vspace{-0pt}
\end{figure}

Beyond the regular functional components of a database system, we refine several key modules within OceanBase Bacchus: \textbf{Shared Log}, \textbf{Metadata Management}, \textbf{Cache Warming}, \textbf{Baseline Compaction}, and \textbf{Garbage Collection}. These modules are organically integrated with clear responsibilities, aiming to reduce storage resources while maximizing the overall system's stability, data consistency, and high availability. Figure~\ref{fig:OBModules} illustrates the information flow between these modules.

Each module serves a distinct purpose:
\begin{itemize}
\item \textbf{Shared Log} manages data consistency, data updates, and recovery operations.
\item \textbf{Metadata Management} handles the fundamental and essential information for all hierarchical modules within the system.
\item \textbf{Cache Warming} ensures fast access to hot data and frequently used metadata in the database.
\item \textbf{Baseline Compaction} is responsible for periodically updating and persisting large volumes of data.
\item \textbf{Garbage Collection} manages the reclamation of old version files and the integration of idle resources within the system.
\end{itemize}

\subsection{Shared Service-Oriented Logging}
\subsubsection{Shared Log} 
\begin{figure}[htp]
    \centering
    \includegraphics[width=0.99\linewidth]{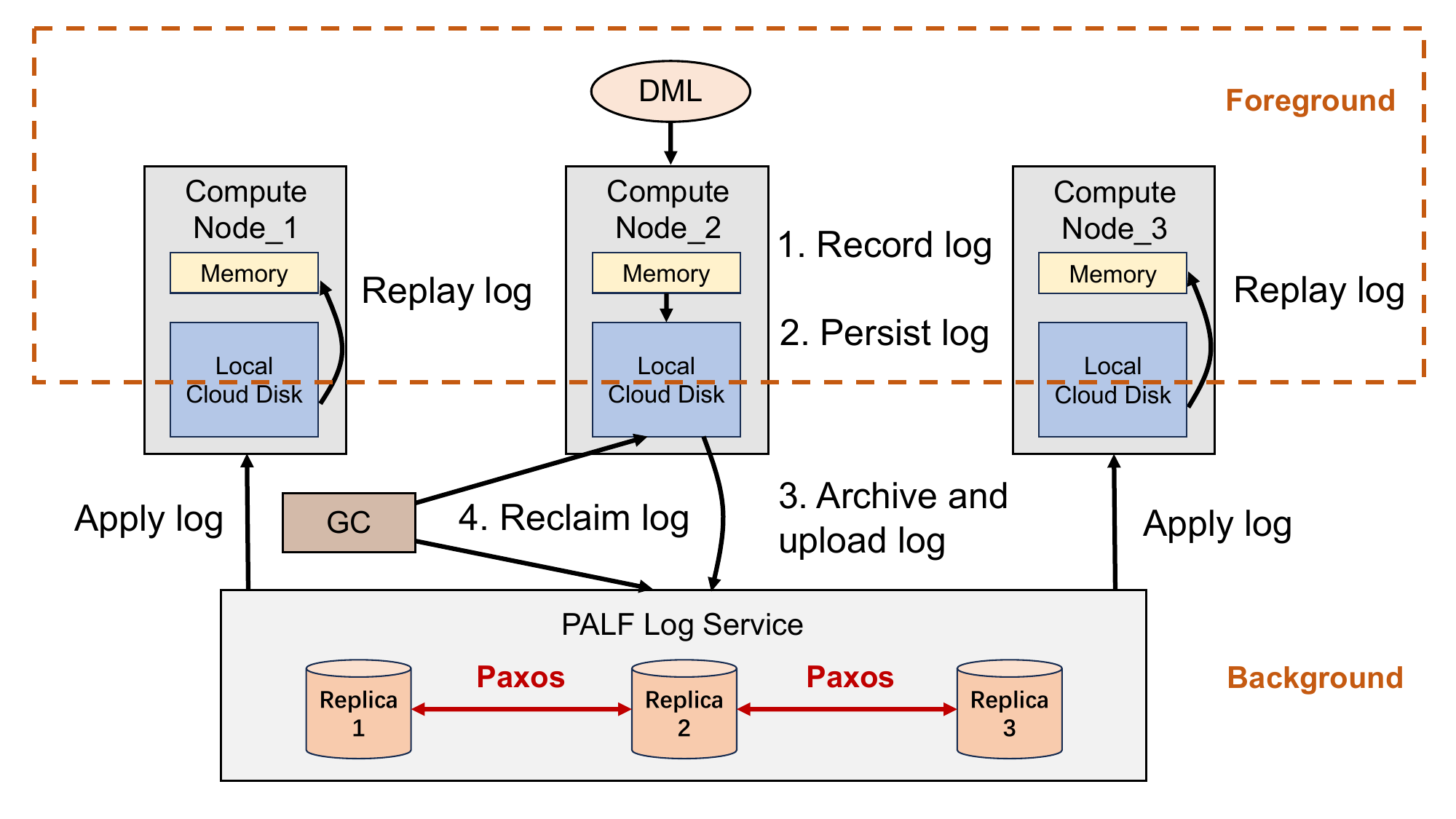}
    \caption{The workflow of Bacchus shared log}
    \label{fig:OBLogWorkFlow}
    \vspace{-0pt}
\end{figure}

Inspired by log and data separation~\cite{10.1145/1376616.1376645}, we adopt a design where multiple partitions share a single log stream, achieving log sharing across clusters. We implement service-oriented logging based on PALF~\cite{10.14778/3685800.3685803}, where logs are persisted and synchronized within a PALF service hosted by LogServer nodes in the shared storage layer. Cloud disks on ECS (Elastic Compute Service) serve as nonvolatile local caches for vital data. Figure~\ref{fig:OBLogWorkFlow} illustrates the process. 

\textbf{Write Coordination and Concurrency Control.} Bacchus employs a sharded write model rather than a true multi-writer system. Data is partitioned into log streams, where each log stream corresponds to a logical data partition (e.g., a set of tablets). Within each log stream, only one node acts as the leader and is responsible for writing CLog entries to the PALF service on the LogServer. This single-writer-per-log-stream design eliminates write conflicts at the log level while allowing multiple RW nodes to operate concurrently on different log streams. The leader election is managed by the database layer using a consensus protocol: when a leader fails, followers elect a new leader for that log stream. This sharding approach ensures that writes to the same logical partition are serialized through a single leader, maintaining consistency without requiring complex distributed locking mechanisms. Multiple RW nodes can coexist in the system, each potentially serving as leader for different log streams, enabling horizontal scaling while preserving write ordering within each partition.

For foreground transactions, CLog entries are first recorded in local cache for low latency. Replicas include a leader and followers, similar to \textit{Coordinator} and \textit{Participant}~\cite{10.5555/647433.723863}. The leader moves historical CLog files to the log service. To meet OLTP performance requirements, Bacchus implements near real-time log archiving for PITR (Point-in-Time Recovery)~\cite{10.1145/3626246.3653382}. The archiving aggregates log writes on cloud disk, uses incremental file uploads via \textit{Append} and \textit{MultiUpload} methods, and supports active flush for faster snapshot generation.

For consistency, we leverage Paxos~\cite{lamport2001paxos} via PALF to ensure strong consistency and durability. User operations generate CLogs, which are synchronized to standby replicas. The protocol guarantees consistency once a majority of replicas persist the log. Minority replica failures do not cause data loss or service interruption. PALF optimizes traditional Paxos through batching and pipelining: multiple log entries are batched into a single consensus round, reducing network round-trips and improving throughput. The pipelining mechanism allows the leader to propose new log entries while previous entries are still being committed, overlapping consensus rounds and achieving lower latency compared to sequential Raft-based replication used in systems like TiDB and CockroachDB. This optimization is particularly beneficial in shared storage architectures where log persistence latency directly impacts transaction commit latency.

Nodes consume logs through a unified mechanism using iterators. Local cloud disk logs are consumed first; if unavailable, service logs are used. After CLog relocation, replicas can reclaim local log files. Garbage collection reclaims shared and local CLog files based on the minimum log replay position and relocation progress.

\subsubsection{Shared Storage Log} Shared Storage Log (SSLog) is a special CLog serving as WAL for metadata. Frequent metadata updates can generate excessive I/O. We introduce region-level SSLog in the log service, using SSLog for metadata while object storage manages data. SSLog stores KV tables, transforming expensive shared storage I/O into cost-effective log-service I/O through aggregation. Like Iceberg~\cite{iceberg} and Delta Lake~\cite{10.14778/3415478.3415560}, we aggregate metadata in SSLog before flushing to reduce I/O overhead.

SSLog benefits scenarios with non-data modifications (ID changes, marker files) and partitions with only metadata changes (empty MemTable dumps, compaction without data, heartbeats). We write metadata updates to SSLog and confirm completion by reading the SSLog tablet. RW nodes write to SSLog instead of sharing metadata directly; RO nodes poll SSLog to update local metadata.

\vspace{-10pt}
\subsection{Metadata Management}

Metadata is hierarchically organized (node, tenant, log stream, tablet) as small files. Bacchus distinguishes the Sys Tenant Cluster (system metadata and control plane) from User Tenant Clusters (user data and application workloads). Each level has node-private and shared types. A metadata service manages metadata, with SSLog stored in the log service. Self-contained metadata with independent lifecycles reduces reliance on global indexes, lowering complexity. We promote block index micro-blocks to memory/cache, reducing random I/O reads.
Different metadata levels use different write strategies. Log stream level and above use write-through (promptly updated to shared storage) due to no multiple versions and lower frequency. 

Tablet level and below use write-back (asynchronous persistence) for multiple versions and real-time performance. Since metadata files are separate, we adjust OceanBase 2PC~\cite{10.14778/3554821.3554830, 10.14778/3611540.3611560}: prepare phase generates metadata files; commit phase updates parent-level files, ensuring consistency.
For shared metadata, concurrent updates by different nodes are managed by a Shared Storage Writer (SSWriter) node per region. All shared tablet metadata modifications go through SSWriter. SSWriter broadcasts changes to other nodes, which trigger local updates upon detection.

\textbf{Table-level Metadata Changes.} Table-level metadata changes (e.g., schema modifications, partition splits, table drops) require careful coordination between the database layer and storage layer. When a table-level operation is initiated, the database layer first writes the metadata change to SSLog, then updates the corresponding metadata files in the metadata service. The change is propagated through SSLog replay: RO nodes poll SSLog and apply metadata changes locally, while RW nodes directly read from SSLog when needed. For atomic table operations (e.g., table drop), we use a two-phase approach: the prepare phase writes the operation intent to SSLog, and the commit phase updates parent-level metadata files. This ensures that table-level changes are visible consistently across all nodes, even if some nodes are temporarily offline. During table-level operations, the database layer maintains transaction isolation by tracking active transactions that reference the table, ensuring that ongoing queries complete before metadata changes take effect.

\subsection{Replication and Migration}
The replication and migration process of OceanBase Bacchus has been adjusted to leverage shared storage. The primary change is that baseline data is directly shared from shared storage, while incremental data and warm data are shared from the distributed cache. Only a small amount of data needs to be copied from the source node to the target node, such as the hottest data in local cache and node-private data/metadata.

In Bacchus, the node replication migration process is as follows: First, a new log stream is created at the target end, without initiating log replay (step 1). Next, an available and suitable source end is selected as the starting point for migration (step 2). Subsequently, the log stream is placed in an offline state, and the corresponding metadata at the target end is updated based on PALF and the log stream information from the source end. Empty shells are created for tablets at the target end; these shells do not contain any data but include the tablet's metadata from the source end (steps 3--4). Then, necessary private information from the source end server is completely copied and transferred (step 5). After that, the log stream is switched to an online state, preparing for subsequent log filtering and replay according to the checkpoint SCN (System Change Number) in the tablet metadata (step 6). Next, tablets copy local cache data in parallel and obtain baseline data from object storage and dumped data from distributed cache. Then, the target executes log replay until the log position grows sufficiently large to ensure that the data is up-to-date (steps 7--8). Finally, after tablets have completed data reference, the member list is updated. The status information during the migration process is cleaned up, updated, and reported (steps 9--10).

\section{Increment and Compaction} \label{sec:OBIncrement}

The compaction process in OceanBase Bacchus is categorized into two primary types: Dumping (Micro/Mini Compaction) and Merging (Minor/Major Compaction \cite{10.1145/1365815.1365816}).

\subsection{Shared Incremental Data}
\begin{figure}[htp]
    \centering
    \includegraphics[trim=0 0 0 0,clip,width=1.0\linewidth]{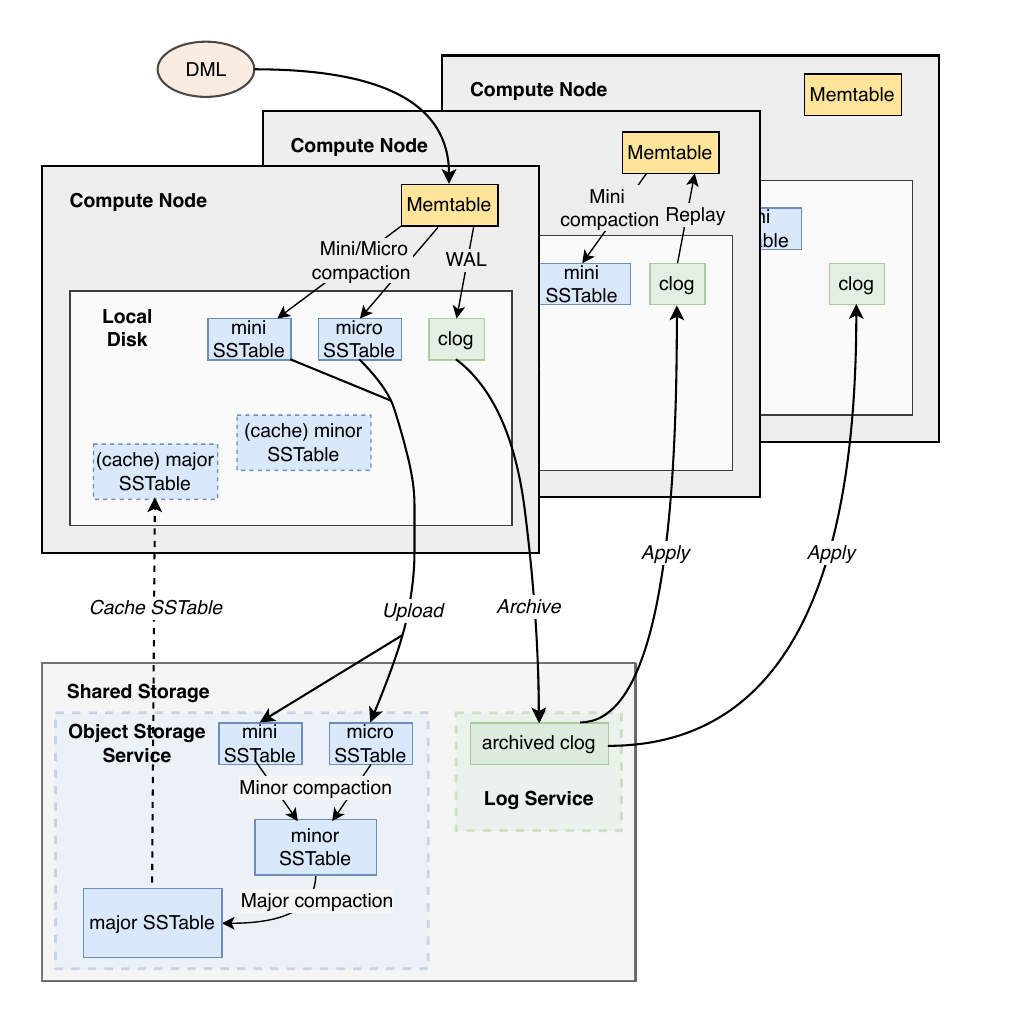}
    \caption{The workflow of Bacchus compaction}
    \label{fig:OBDataWorkFlow}
    \vspace{-0pt}
\end{figure}

It is essential to ensure that incremental data in the local cache is written to OSS promptly. Bacchus implements a fast dumping strategy for incremental data. The collaborative processing workflow for shared CLog and incremental data is illustrated in Figure~\ref{fig:OBDataWorkFlow}. User data is first written into the MemTable in memory, and transaction persistence is achieved through CLog. When the memory usage of MemTable reaches a certain threshold, MemTable is frozen to generate a mini SSTable by mini compaction, thus releasing memory resources, which we call a ``checkpoint" of logging. Since the process of crash recovery and replica addition requires replaying all data corresponding to the MemTable, we introduce micro compaction (and micro SSTables, a type of smaller SSTables) to further reduce the cost of elastic scaling. By generating micro SSTables promptly before freezing MemTable, we can more quickly advance the checkpoint position of the log, thereby accelerating crash recovery and replica loading. Since each compute node has different timing and urgency for releasing memory, MemTable generates the corresponding SSTable (mini/micro SSTable) as quickly as possible and then frees up space; otherwise, it will affect service capability in OLTP scenarios. Therefore, we first place the generated micro SSTables and mini SSTables in the local disk caches of compute nodes, and then a specific replica uploads the cached SSTables to object storage in the background. Next, minor compaction~\cite{10.1145/1365815.1365816} merges multiple micro/mini/minor SSTables in shared storage into a single minor SSTable. 

During the minor compaction process, macro-block-level reuse is performed to control the overall write amplification of the system. After minor compaction, GC (garbage collection) can delete the input SSTables in shared storage to save storage space. Through these stages, we have implemented a faster process of dumping and persisting data. It should be noted that there may be multiple demands for shared data within the cluster in a short period of time, and the object storage interface lacks mutual exclusion capabilities. To address this potential issue, we rely on the election capability of the compute nodes, allowing the leader of each log stream to select a replica with relatively lower load to act as Shared Storage Writer (SSWriter) for that log stream in shared storage. Within the validity period of the given lease of SSWriter, only this replica is permitted to execute object storage write tasks for all tablets under the log stream.

\subsection{Major Compaction}

We design Major Compaction (MC) according to the characteristics of object storage. When the number of mini compactions (dumping operations) exceeds a predefined threshold, or during the off-peak business period each day, the system merges the baseline SSTable (Major SSTable) with subsequently dumped incremental SSTables (Mini/Minor SSTables) into a new Major SSTable. This process, termed major compaction, is divided into 7 phases. Our self-developed daily compaction achieves the separation of front-end and back-end operations. The major compaction process takes place primarily within the shared storage layer, independent of the services provided by the database layer, with compute nodes being agnostic to the specific process. Algorithm~\ref{alg:OBCompaction} and \ref{alg:OBCompaction2} illustrate the main process of compaction.

\begin{enumerate}
\item The Root Service (RS) at the database layer initiates the daily Major Compaction (MC) process, and the User Tenant Cluster's compute node detects the pending merge task.
\item The leader of compute nodes schedules the merging of tablets, distributes compaction tasks of tablets, and writes the tasks into the metadata service of the shared storage layer.
\item The node in the shared storage layer responsible for executing MC detects the pending merge task, performs the major compaction, merges minor/major SSTables into one major SSTable, and stores the resulting SSTable in the object storage service.
\item After the compaction is completed, the execution node updates the tablet metadata to the metadata service in the shared storage layer and marks the merge task as completed.
\item The User Tenant Cluster's compute nodes detect the completion of the compaction through replaying the SSLog in the metadata service, reference the new data, warm it up locally, and then report the Cyclic Redundancy Check (CRC) \cite{4066263} checksum of the replica itself to the internal table.
\item The RS detects the completion of the replica warm-up through the internal table, performs checksum verification between the primary table and index table, and verifies cross-region replica consistency by replica checksum to guarantee data consistency between primary/index tables and ensure that business data is consistent across different replicas.
\item Upon successful verification by the RS, the daily compaction process is completed.
\end{enumerate}

\begin{algorithm}
\caption{Major Compaction Algorithm in Database Layer}
\label{alg:OBCompaction}
\begin{algorithmic}[1]
\REQUIRE A root service \(RS\). A metadata service \(MS\). A leader replica \( l \) and a set of replicas \( F: \{f_1, f_2, \ldots, f_n\} \). Each replica \( f_i \in F \) corresponds to cached SSTable(s) \( D_i \). A checksum algorithm \( \textbf{crc}: D \rightarrow \mathbb{R}, D \in SSTable.\)
\ENSURE A Major Compaction (MC) adapted to the characteristics of shared storage.
\STATE \(RS\) launches MC. \( l \) receives MC. 
\STATE  \( l \) schedules tablets and \( MS \gets PrepareMC(l) \)
\STATE Call Algorithm~\ref{alg:OBCompaction2} and get Major SSTable \( D_{\text{baseline}} \)
\STATE \( checksum_{\text{exec}} \gets \textbf{crc}(D_{\text{baseline}}) \)
\FOR{$f_i \in F$}
    \STATE \( D_i \gets PreheatCache(f_i, D_{\text{baseline}}), \   checksum_{i} \gets \textbf{crc}(D_i) \)
    \IF{$checksum_{i} \neq checksum_{\text{exec}}$} 
    \STATE Retry
    \ENDIF
\ENDFOR
\STATE \(RS\) checks \( \textbf{crc}(Primary Table) = \textbf{crc}(Index Table) \)
\end{algorithmic}
\end{algorithm}
\begin{algorithm}
\caption{Major Compaction Algorithm in Shared Storage Layer}
\label{alg:OBCompaction2}
\begin{algorithmic}[1]
\REQUIRE A metadata service \(MS\). A node \( N \) in the shared storage layer that executes a major compaction (MC).
\STATE \( D_{\text{increment}} \gets MinorSSTable,   D_{\text{baseline}} \gets MajorSSTable\)
\STATE \( N \gets SelectMCNode(MS) \)
\STATE \(D_{\text{baseline}} \gets MC(N, D_{\text{baseline}}, D_{\text{increment}})\)
\STATE \( MS \gets SendMCMessage(N) \)
\RETURN A major SSTable \(D_{\text{baseline}}\) in shared storage.
\end{algorithmic}
\end{algorithm}


\subsection{Compaction Offloading}
We have completed the offloading of minor/major compaction tasks from the database layer to the shared storage layer, where compaction tasks are executed by machines within the shared storage layer. This achieves separation between foreground and background operations. However, since major compaction is a compute-intensive task, the shared storage layer can also offload compaction tasks to compute nodes, further achieving separation of storage and compute and increasing the utilization of idle computational resources.

With shared storage supporting metadata sharing, transaction tables, and incremental data, we can offload many computational requirements to idle machines, utilizing the shared data to perform computational tasks. The compaction offloading process proceeds as follows. First, we choose a certain machine to perform compaction offloading tasks (step 1). Then, a log stream is created that carries information about tablets, data, and transactions for compaction (step 2). The log stream leader sets the machine as an SSWriter. Within the scope of the lease, only SSWriter is allowed to perform write tasks on shared storage (step 3). Next, it loads tablets to obtain the SSTable list and triggers compaction and metadata updates (steps 4--5). The newly generated data is preloaded to local caches on all RW/RO nodes (step 6). After verifying the checksum, the machine is released to the underutilized resource pool.

\section{Caching and Preheating} \label{sec:caching}

\subsection{General Cache and Cache Warming}

The local cache space is divided into three parts: tablet metadata cache, dump and temporary file cache, and micro-block cache. The first two types of data, due to their relatively small file sizes, are generally fully cached. The last type, the micro-block cache, employs the ARC (Adaptive Replacement Cache) algorithm \cite{10.5555/1973355.1973364}. The ARC algorithm is an adaptive cache replacement algorithm that dynamically adjusts the caching strategy and leverages a history-driven adaptive mechanism. By combining the advantages of LRU (Least Recently Used) and LFU (Least Frequently Used), it achieves near-optimal caching efficiency under complex workloads and has become one of the mainstream industrial alternatives to LRU. In OceanBase Bacchus, the data blocks for LRU and LFU are stored in the local cache, while the ghost cache blocks for LRU and LFU are stored in object storage.

To ensure the stability of throughput and latency under various operating conditions, we provide multiple data preheating methods for different scenarios:

\noindent$\bullet\ $ \textbf{Baseline Switching Preheating}. Bacchus shares baselines along with shared logs. During a major version switch, we warm up the new version's baseline by loading the new version's hot macro-blocks into the cache to minimize performance fluctuations.

\noindent$\bullet\ $\textbf{Leader/Follower Replica Preheating}. When switching between leader and follower replicas, to reduce latency fluctuations for user requests, the leader analyzes the access sequence based on the log stream and periodically synchronizes this sequence with followers. Followers then warm up their local micro-block cache according to the access sequence to ensure that the caches of the leader and follower replicas are similar.

\noindent$\bullet\ $ \textbf{Replication Migration Preheating}. During replication migration, incremental data is shared from the Shared Block Cache Service to the target, with the baseline shared from shared storage. Subsequently, hot data is copied from the source to the target on a macro-block basis, and the target writes these data into its local micro-block cache.

\noindent $\bullet\ $\textbf{Cloud Disk Scaling Preheating}. Considering that local cache disks may be scaled up or down, the LRU and LFU lists of ARC need to transfer items with their respective ghost caches accordingly. During scaling up, the list size increases, and items are moved from the ghost cache to the list. During scaling down, items from the list are moved into the ghost cache, and items on the original ghost list are eliminated as needed.

\subsection{Distributed Shared Cache}
Due to the high latency of object storage (100ms+), shared storage with only local caching (ms+) is clearly insufficient to support core OLTP business. Additionally, the bandwidth of object storage is limited. For frequently modified data, employing a distributed cache can alleviate the bandwidth pressure on object storage. 

\begin{figure}[htp]
    \centering
    \includegraphics[width=0.9\linewidth]{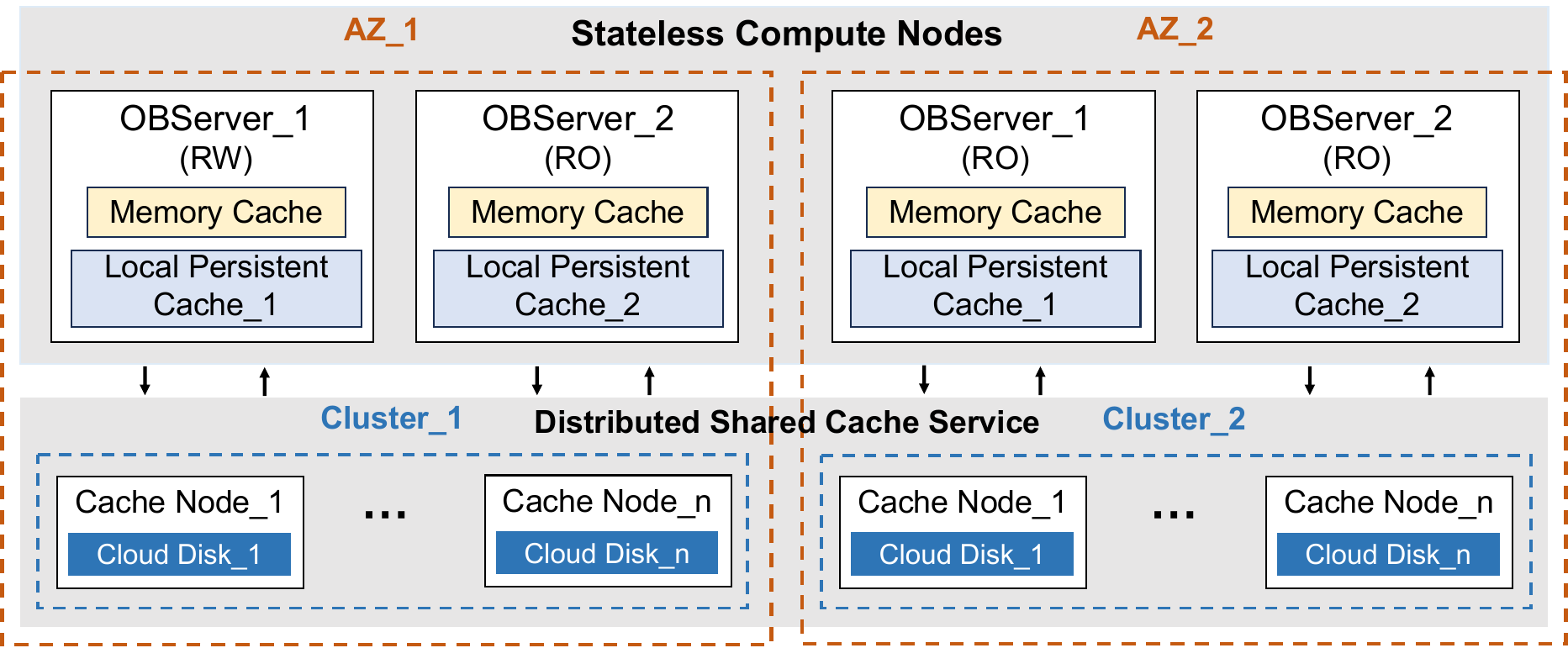}
    \caption{Three-level cache hierarchy}
    \label{fig:OBSharedCache}
    \vspace{-0pt}
\end{figure}

Bacchus draws on the strengths of various solutions, including AWS S3 Express One Zone \cite{s3} (which can be used as a distributed cache at the macro-block level in scenarios where cross-data center operations are not required) and Microsoft Socrates's two-layer cache architecture \cite{10.1145/3299869.3314047} (one layer caches the hottest pages as a sparse cache, the other layer is a dense cache). We develop our own distributed persistent caching service, the Shared Block Cache Service, as displayed in Figure~\ref{fig:OBSharedCache}. It is implemented by BlockServer nodes that store and serve macro-blocks. The service employs a three-tier cache architecture: memory, local cache on compute nodes, and distributed cache, which store the hottest, second hottest, and warm data respectively. The storage granularity also increases progressively: from micro-blocks in local cache to macro-blocks in distributed cache. During cache preheating, the three-tier caches are warmed up in sequence.

The distributed cache is a read-only cache, independent of Bacchus clusters. Each Availability Zone (AZ) deploys a set of distributed caches, and RO/RW compute nodes within the same AZ can share data in the distributed cache, removing redundant copies in nodes and saving cache resources. Bacchus implements a service-oriented distributed cache system (the Shared Block Cache Service), where BlockServer nodes in the service use local cloud disks as cache disks. Relying on the high availability of cloud disk data, data will not be lost after crashes, and the cache service can be quickly restarted. Moreover, with the use of shared caches, compute nodes do not need to copy much data during scaling or migration; they only need to consider a small amount of local metadata/data on nodes. Additionally, scaling up or down of cache nodes has no impact on compute nodes. Although cache nodes consume additional ECS resources, they are mainly I/O-intensive. With the enhanced fast elasticity capability of the Bacchus server, the overall computing cost is still reduced.

Compared with caching services from other cloud vendors, since OceanBase is a multi-cloud native database, its self-developed distributed shared caching framework is tailor-made for OceanBase and has a better understanding of the database's access model. Therefore, it achieves better performance and stability.

\subsection{Concurrency Control in Multi-level Caching} 

The three-level cache architecture (memory, local cache, distributed cache) requires careful concurrency control to maintain consistency. When data is modified in the database layer, the system uses a write-through policy for the memory cache and a write-back policy for local and distributed caches. Cache invalidation is coordinated through SSLog: when metadata changes occur (e.g., SSTable updates after compaction), the change is logged in SSLog, and all nodes invalidate corresponding cache entries upon replaying the log. For distributed cache, we employ a version-based invalidation scheme where each cache entry is tagged with a version number. When a node updates data, it increments the version and writes the change to SSLog. Other nodes detect version mismatches when accessing cached data and refresh from shared storage. This ensures that stale cached data is never served, even in the presence of concurrent updates across multiple nodes.

\section{Garbage Collection} \label{sec:GC}

Garbage collection in Bacchus must coordinate with the database layer to ensure correctness while reclaiming obsolete data. The system tracks data versions using SCN and maintains visibility information to determine when data can be safely reclaimed.

\subsection{Global Garbage Collection Coordination} 

Global GC in Bacchus reclaims shared CLog files and obsolete SSTables across multiple compute nodes, requiring careful coordination to prevent concurrent deletion conflicts and ensure consistency. We employ a lease-based distributed coordination mechanism similar to SSWriter but specifically designed for GC tasks. Each log stream designates a \textit{GC Coordinator} node through leader election; this node is responsible for orchestrating GC operations for that log stream's data. The GC Coordinator acquires a time-bound lease (typically 30-60 seconds) from the metadata service, recorded in SSLog to ensure visibility across all nodes. During the lease period, only the GC Coordinator is permitted to delete shared storage objects (CLog files and SSTables) for its assigned log stream, preventing race conditions where multiple nodes might attempt to delete the same file concurrently.

The GC coordination process operates as follows: (1) \textit{Lease Acquisition}: The GC Coordinator periodically renews its lease by writing a lease record to SSLog with a timestamp and expiration. If a GC Coordinator fails, its lease expires, and a new coordinator is elected. (2) \textit{Safe Reclamation Point Calculation}: Before deleting any data, the GC Coordinator queries the metadata service to obtain the global min\_read\_scn and the minimum log replay position across all nodes. It only marks data for deletion if it is older than these safety thresholds. (3) \textit{Atomic Deletion}: The GC Coordinator first writes a deletion intent to SSLog, listing the files to be deleted. After a short grace period (to allow other nodes to detect the intent), it proceeds with actual deletion. If the deletion fails partially, the intent record enables recovery and prevents orphaned references. (4) \textit{Metadata Synchronization}: After successful deletion, the GC Coordinator updates metadata files to remove references to deleted objects, and this change propagates to all nodes through SSLog replay. This ensures that all nodes maintain consistent views of available data files.

To handle failures and prevent GC stalls, we implement lease renewal with exponential backoff: if a GC Coordinator cannot renew its lease (e.g., due to network partition), it stops GC operations and releases its lease. Other nodes detect the expired lease and elect a new coordinator. Additionally, we use a two-phase deletion protocol: the prepare phase marks files as "pending deletion" in metadata, and the commit phase performs actual deletion only after confirming that no active transactions reference these files. This mechanism ensures that GC operations are safe, consistent, and resilient to node failures while maintaining high availability of the GC service.

\subsection{Cross-Partition Transaction Coordination} 

While each log stream has a single writer (leader), transactions may span multiple log streams, requiring coordination across partitions. Bacchus uses a distributed transaction protocol (OceanBase 2PC \cite{10.14778/3554821.3554830}) where a coordinator node manages the transaction across multiple log stream leaders. The coordinator first collects prepare votes from all participating log stream leaders, each of which writes a prepare log entry to its respective PALF log stream. Once all participants vote to commit, the coordinator writes a commit log entry, and all participants write commit entries to their log streams. This ensures atomicity across partitions: either all log streams commit the transaction or all abort. The PALF service's consensus mechanism guarantees that commit decisions are durable and consistent across all replicas. Since each log stream maintains its own write ordering through its leader, and cross-partition coordination is handled at the transaction level, the system achieves both partition-level parallelism and global transaction consistency. Distributed deadlock detection and resolution in this setting is addressed by the LCL~\cite{yang2023lcl} and LCL+~\cite{yang2025lcl+} algorithms.

\subsection{Handling Long-running Transactions} 

Long-running open transactions~\cite{fang2025malt} pose a challenge for GC, as they may require access to historical data versions that would otherwise be eligible for reclamation. Bacchus addresses this through a multi-version visibility mechanism. Each transaction is assigned a read SCN at its start, which determines the data versions it can access. The database layer maintains a global minimum active read SCN (min\_read\_scn) across all nodes, representing the oldest active transaction's read point. GC only reclaims data versions older than min\_read\_scn, ensuring that no active transaction loses access to required data. When a transaction exceeds a configurable timeout threshold, the database layer either aborts it or promotes its read SCN to prevent indefinite data retention. This mechanism is coordinated through the metadata service: each compute node periodically reports its local minimum read SCN, and the metadata service aggregates these to compute the global min\_read\_scn, which is then broadcast to all nodes for GC decisions.

\section{Performance Evaluation} \label{sec:eval}

\subsection{Experimental Configuration}

We selected the following databases for evaluation: Bacchus (OceanBase 4.4.x), Paetica (OceanBase 4.0), HBase~\cite{hbase}, and StarRocks~\cite{starrocks}. The experiments cover both \textbf{transactional (OLTP)} and \textbf{analytical (OLAP)} workloads: OLTP is evaluated with SysBench~\cite{sysbench} and write/read benchmarks (e.g., vs. HBase, Aurora, TDSQL, PolarDB); OLAP is evaluated with TPC-H~\cite{tpch}, TPC-DS~\cite{tpcds}, and ClickBench~\cite{clickbench} against StarRocks, since TPC-H and TPC-DS are standard benchmarks for analytical and decision-support queries.

Single-node experiments in \textsection \ref{sec:obvshbase} are performed on ecs.r7.2xlarge instances in Alibaba Cloud, with ecs.g7.xlarge instances used for log storage. Both OceanBase and PolarDB in \textsection \ref{sec:obvspolardb} are deployed in Alibaba Cloud with 32 cores. Single-node experiments in \textsection \ref{sec:obvsaurora} are performed on db.r5.xlarge instances in AWS.
Single-node experiments in \textsection \ref{sec:obvsstarrocks} are conducted on two hardware configurations: (1) 32-core Intel(R) Xeon(R) Platinum 8575C CPU with 256GB DRAM for TPC-H 100GB, and 48-core Intel(R) Xeon(R) Platinum 8575C CPU with 384GB DRAM for TPC-H 1TB; (2) 16-core Intel(R) Xeon(R) Platinum 8369B CPU with 64GB DRAM for TPC-DS and ClickBench. In \textsection \ref{sec:obmulticloud}, we deploy OceanBase in Alibaba Cloud with 32-core Intel(R) Xeon(R) Platinum 8575C.

\subsection{Performance Comparison}

\subsubsection{OceanBase 4.4.x vs. HBase 1.4.7} \label{sec:obvshbase}
\begin{figure}[htp]
  \centering
  \includegraphics[width=0.9\linewidth]{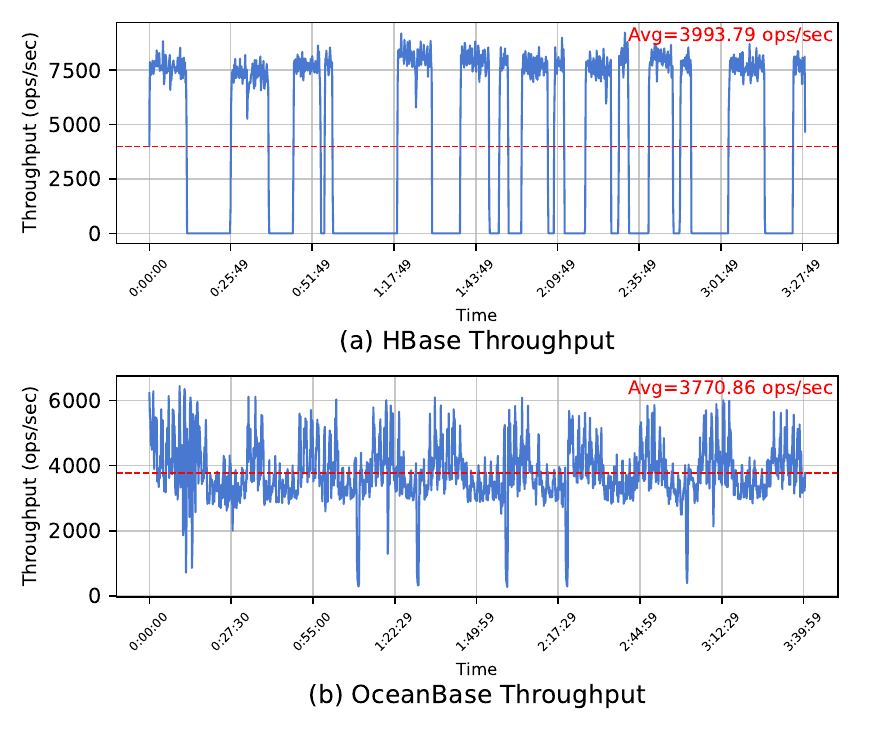}
  \caption{Write throughput in OceanBase and HBase}
  \label{fig:OBHbase}
  \vspace{-0pt}
\end{figure}

In this experiment, we fixed the dataset size at 500 GB and compared the write throughput of OceanBase Bacchus and HBase over a sustained period. As shown in Figure~\ref{fig:OBHbase}, although the average write throughput of the two systems was roughly equivalent (3,770.86 vs. 3,993.79 ops/s), HBase unexpectedly suffered from intermittent and continuous write-drop-to-zero events. This was most likely caused by foreground write traffic rapidly reaching its peak while flush and dump operations could not keep pace, triggering foreground write blocking. While raising the frequency of background flushes and compactions would be effective, it reduces the peak write performance in turn. Bacchus, although exhibiting a lower peak throughput than HBase, avoids write stalls completely through its self-developed fast dump strategy, thereby sustaining an overall write throughput comparable to HBase.

To further investigate Bacchus's performance advantages, we conducted additional comparative evaluations against multiple HBase variants. Our experimental setup included Alibaba Cloud Lindorm HBase (version 2.0), Tencent Cloud HBase (EMR Serverless version), and previous OceanBase (version 4.2.5) as baselines. All systems were deployed with two nodes, except for Tencent HBase, which was configured with three nodes due to vendor constraints. All experiments were conducted under uniform hardware conditions (16 CPU cores, 32 GB RAM, and over 1 TB of storage) with a concurrency of 200, and consisted of an empty-table write stress test (PUT) and a random read test (GET) with 500 million records.

\begin{figure}[htp]
  \centering
  \includegraphics[width=0.95\linewidth]{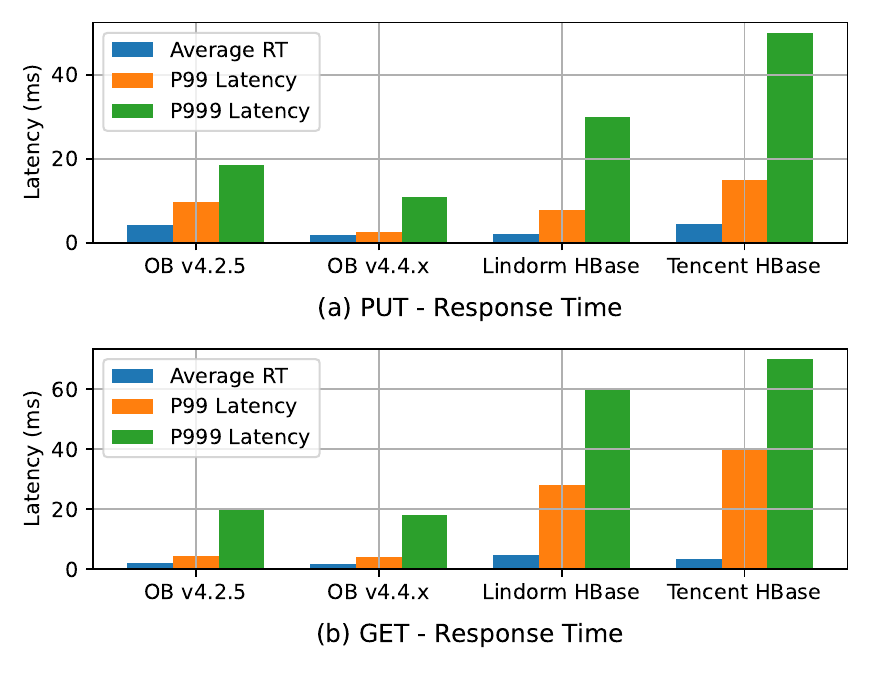}
  \caption{Response time in OceanBase and HBase}
  \label{fig:OBHbase2}
  \vspace{-0pt}
\end{figure}

In PUT experiments, Bacchus achieved the highest performance in both total TPS and per-node average TPS (Total: 51,612 TPS; Avg: 25,806 TPS). Its mean and percentile response time metrics were significantly better than the other three systems, with mean latency being only 41.8\% of that observed in the legacy OceanBase. In GET experiments, Bacchus's throughput on two nodes was marginally lower than that of the three-node Tencent HBase, but its per-node average QPS still led by a wide margin at 28,208 QPS, and response performance remained the highest. Detailed results can be found in Table~\ref{tab:OBHbase2} and Figure~\ref{fig:OBHbase2}.

\begin{table}[htp]
\caption{TPS/QPS in OceanBase and HBase}
\small
\begin{tabular}{|c|cc|cc|}
\hline
{}                                                        & \multicolumn{2}{c|}{{\textbf{PUT}}}                                                                   & \multicolumn{2}{c|}{{\textbf{GET}}}                                                                   \\ \cline{2-5} 
\multirow{-2}{*}{\textbf{Product}}                               & \multicolumn{1}{c|}{{\textbf{Total}}}              & {\textbf{Average}} & \multicolumn{1}{c|}{{\textbf{Total}}}              & {\textbf{Average}} \\ \hline
{\textbf{OB v4.2.5}}                 & \multicolumn{1}{c|}{{45,332}}                          & {22,666}                & \multicolumn{1}{c|}{{48,800}}                          & {24,400}                \\ \hline
 
{\textbf{OB v4.4.x (Bacchus)}} & \multicolumn{1}{c|}{{\textbf{51,612}}} & {\textbf{25,806}}               & \multicolumn{1}{c|}{{56,416}}  & {\textbf{28,208}}               \\ \hline
{\textbf{Lindorm HBase}}                        & \multicolumn{1}{c|}{{44,028}}                          & {22,014}                & \multicolumn{1}{c|}{{16,000}}                          & {8,000}                 \\ \hline
 
{\textbf{Tencent HBase}}                        & \multicolumn{1}{c|}{{40,770}}  & {13,590}                & \multicolumn{1}{c|}{\textbf{60,600}} & {20,200}                \\ \hline
\end{tabular}
\label{tab:OBHbase2}
\end{table}

\subsubsection{OceanBase 4.4.x vs. PolarDB MySQL 8.0.1} \label{sec:obvspolardb}

\begin{figure}[htp]
  \centering
  \includegraphics[width=0.9\linewidth]{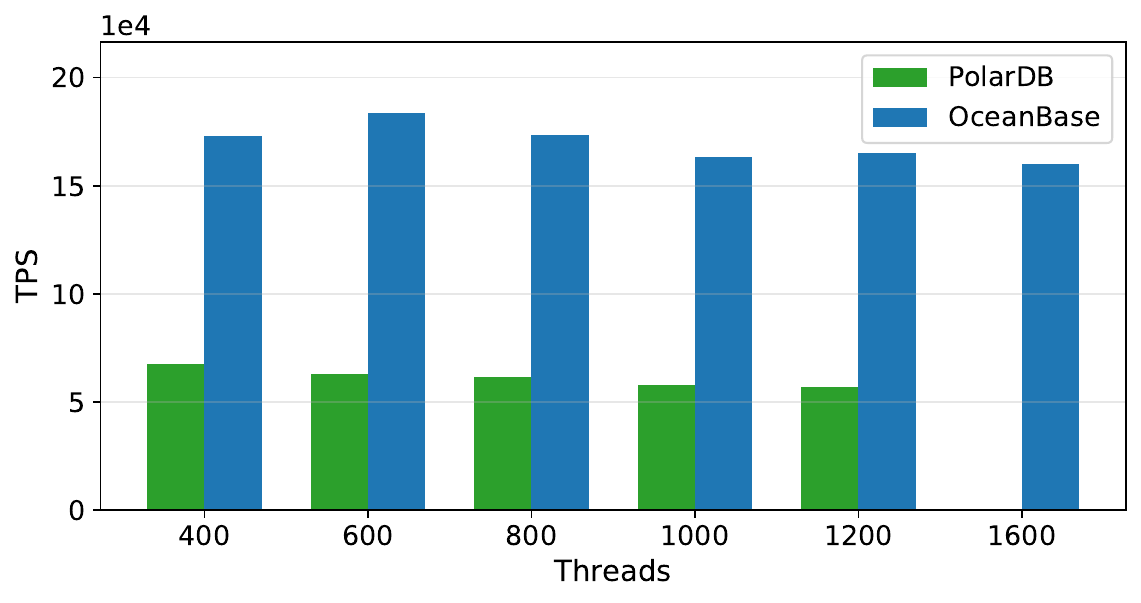}
  \caption{Average Throughput in OceanBase and PolarDB}
  \label{fig:OBPolarDB}
  \vspace{-0pt}
\end{figure}

In a 32 vCPU, 128 GB RAM environment, we loaded a unified dataset of 30 tables with 1,000,000 rows each. We benchmarked OceanBase and PolarDB under four workload types: \textit{Point Select}, \textit{Read-Only}, \textit{Write-Only}, and \textit{Read–Write}, and reported the aggregate average throughput across these scenarios. As shown in Figure~\ref{fig:OBPolarDB},
OceanBase achieved notably higher throughput (TPS) than PolarDB, with only a slight increase in latency. As concurrency (number of threads) increased, OceanBase maintained stable, high throughput. At 1,600 threads, PolarDB failed, while OceanBase continued to serve the workload stably. The results demonstrated that OceanBase offered higher throughput and higher stability than cutting-edge databases in the tested scenarios.

\subsubsection{OceanBase 4.4.x vs. Aurora 3.08} \label{sec:obvsaurora}

\begin{figure}[htp]
  \centering
  \includegraphics[width=1.0\linewidth]{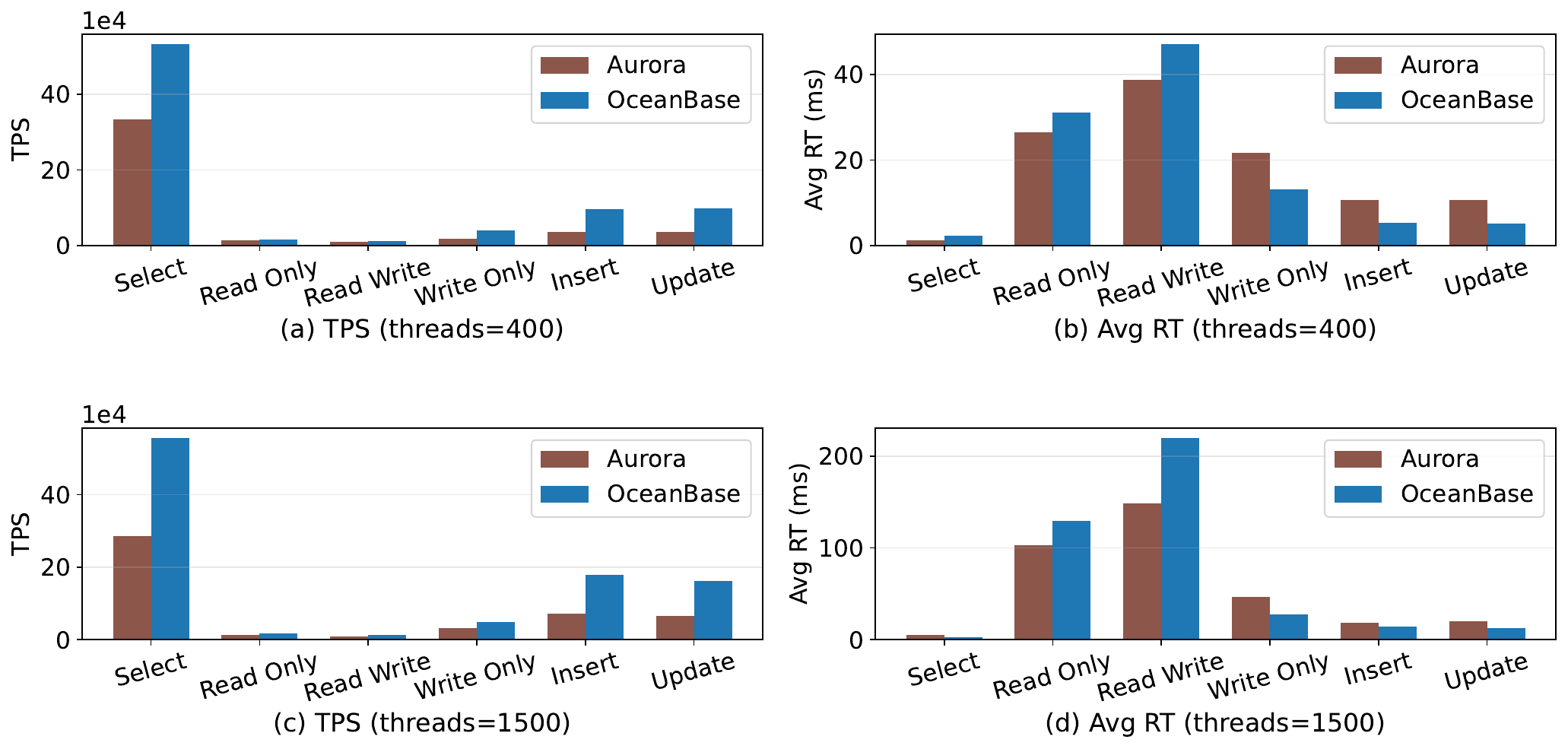}
  \caption{Throughput and Latency in OceanBase and Aurora}
  \label{fig:OBAurora}
  \vspace{-0pt}
\end{figure}

Figure~\ref{fig:OBAurora} summarizes the results on a 32C128G configuration using 30 non-partitioned tables with 500K rows. OceanBase and Aurora exhibited distinct performance characteristics in terms of throughput and latency. Across \textit{Point Select}, \textit{Write-Only}, \textit{Insert}, and \textit{Update} workloads, OceanBase achieved significantly higher TPS under high concurrency (e.g., 400–1500 threads), with throughput exceeding Aurora by 2× in high-thread \textit{Insert} and \textit{Update} scenarios. However, OceanBase demonstrated steeper latency escalation at elevated thread counts (e.g., \textit{Read-Only} RT reaching hundreds of milliseconds at 1500 threads), whereas Aurora maintained more stable RT in read-intensive workloads. Overall, under identical data scale and default configurations, OceanBase on 32C128G favored high-throughput scenarios, while Aurora delivered more balanced latency for read workloads.

\subsubsection{OceanBase 4.4.x vs. TDSQL MySQL 8.0} \label{sec:obvstdsql}

\begin{figure}[htp]
  \centering
  \includegraphics[width=1.0\linewidth]{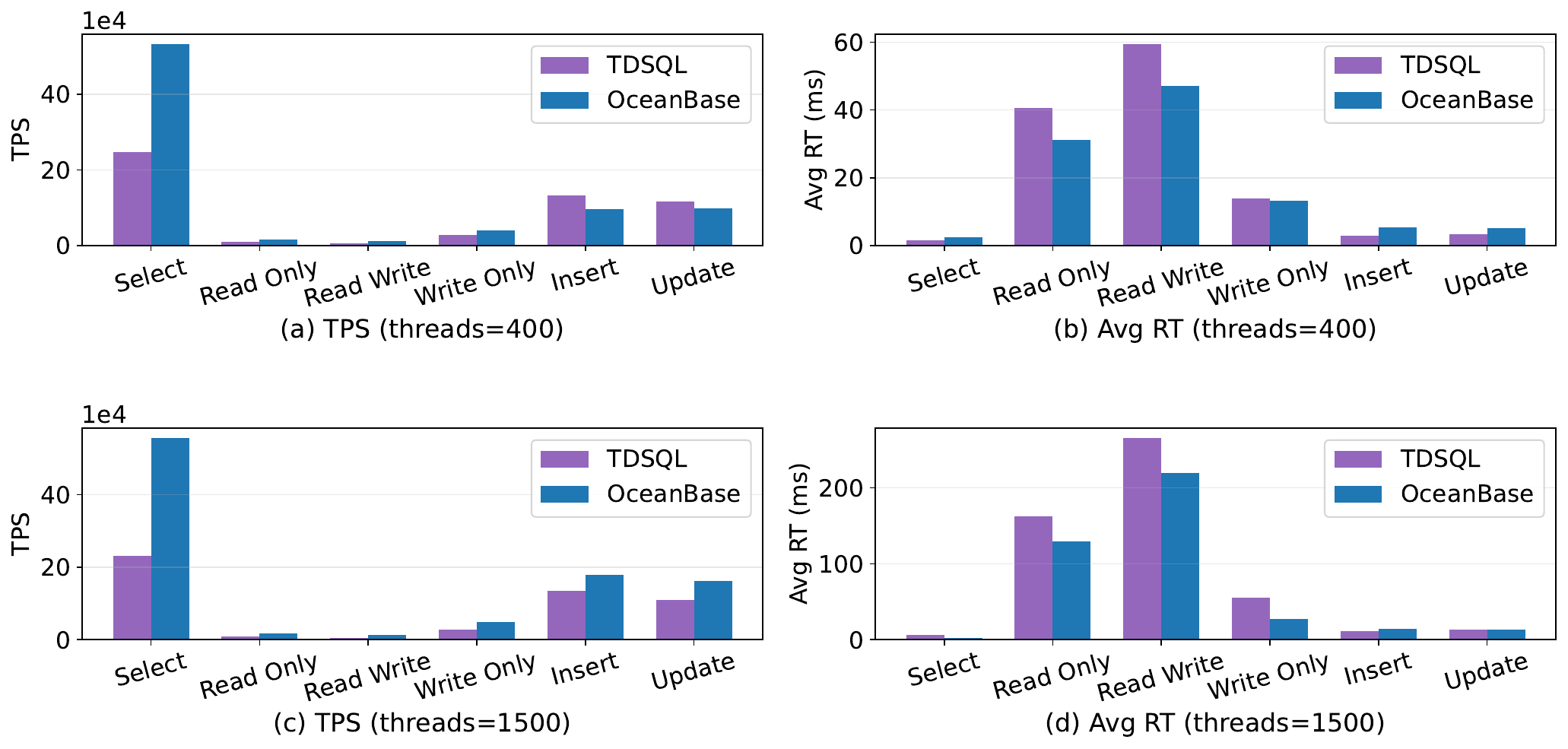}
  \caption{Throughput and Latency in OceanBase and TDSQL}
  \label{fig:OBTDSQL}
  \vspace{-0pt}
\end{figure}

Based on the data shown in Figure~\ref{fig:OBTDSQL}, under a 32-core/128GB memory configuration with 30 non-partitioned tables and approximately 500,000 rows, OceanBase and TDSQL~\cite{10.14778/3352063.3352122} excelled in different aspects. In read-only workloads (\textit{Select} and \textit{Read-Only} scenarios), OceanBase achieved higher TPS at both 400 and 1,500 threads with more than twice the TPS of TDSQL in \textit{Select}. As concurrency increased, in mixed read-write and write-intensive workloads (\textit{Read-Write}, \textit{Write-Only}, \textit{Insert}, and \textit{Update} scenarios), OceanBase's peak throughput generally surpassed that of TDSQL (for instance, OceanBase came from behind at 1,500 threads in \textit{Insert} and \textit{Update}). Regarding average Response Time (RT) under high concurrency, both databases had advantages in different scenarios; however, TDSQL's average RT was slightly higher than OceanBase's across 6 scenarios. In summary, OceanBase demonstrated consistently high throughput performance, while TDSQL exhibited more controllable latency in certain high-concurrency scenarios.

\subsubsection{OceanBase 4.4.x vs. StarRocks 3.3.13} \label{sec:obvsstarrocks}

\begin{figure*}[htb]
\centering
\setlength{\floatsep}{5pt plus 3pt minus 2pt}
\setlength{\textfloatsep}{5pt plus 3pt minus 2pt}
\setlength{\intextsep}{5pt plus 3pt minus 2pt}
\begin{subfigure}[t]{0.49\textwidth}
  \centering
  \includegraphics[width=\linewidth]{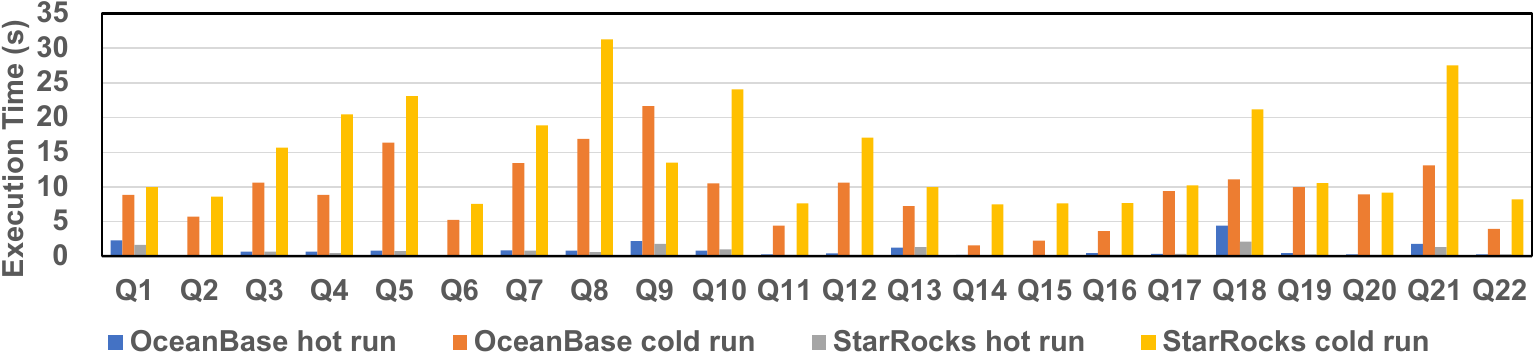}
  \caption{TPC-H 100GB}
  \label{figOBStarRocksTPCH100GB}
\end{subfigure}\hfill
\begin{subfigure}[t]{0.49\textwidth}
  \centering
  \includegraphics[width=\linewidth]{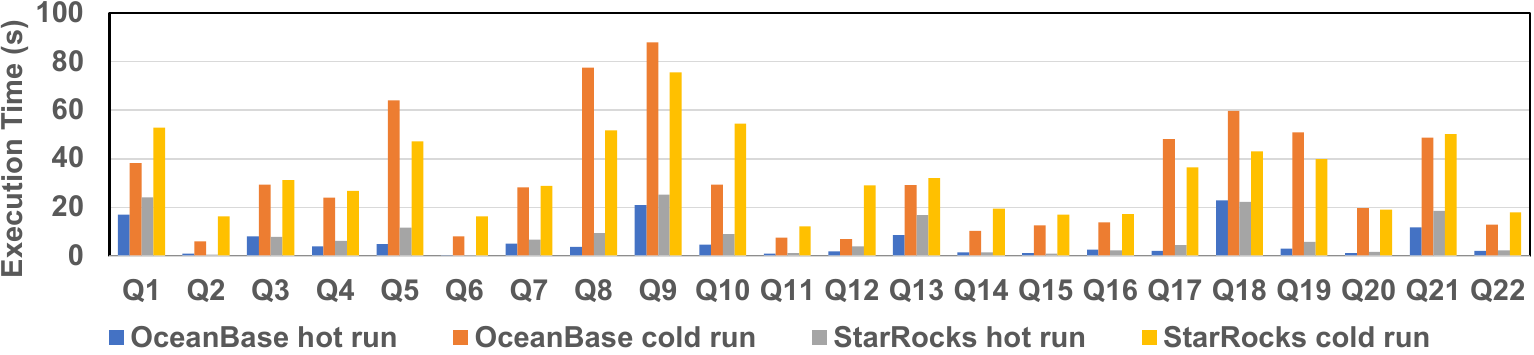}
  \caption{TPC-H 1TB}
  \label{figOBStarRocksTPCH1TB}
\end{subfigure}
\caption{Query performance of OceanBase and StarRocks with different benchmarks}
\label{fig:OBStarRocksTPCH}
\end{figure*}

\begin{figure*}[htp]
  \centering
  \includegraphics[width=0.95\linewidth]{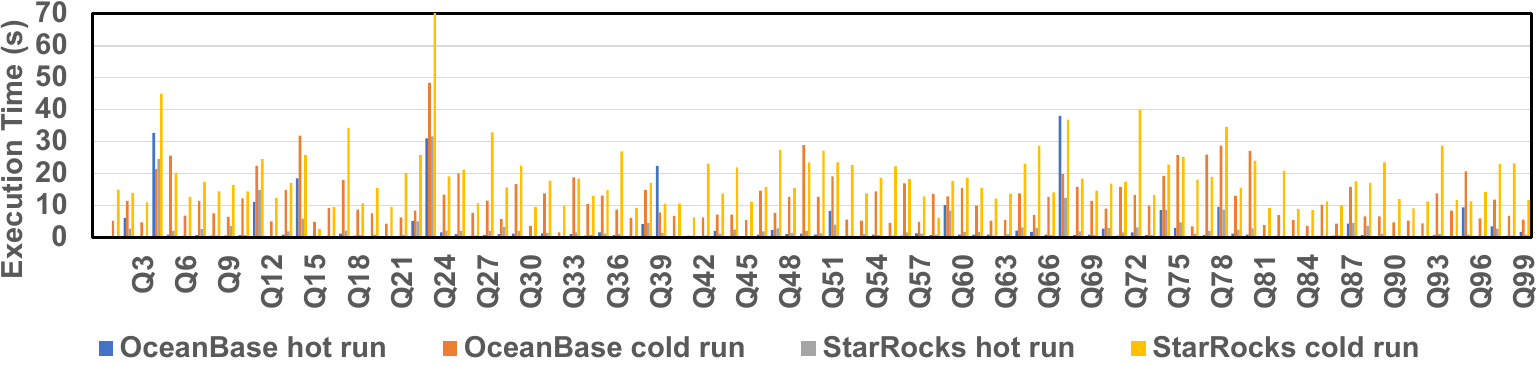}
  \caption{Query performance of OceanBase and StarRocks with TPC-DS 100GB}
  \label{fig:OBStarRocksTPCDS100GB}
  \vspace{-0pt}
\end{figure*}

\begin{figure*}[htp]
  \centering
  \includegraphics[width=0.95\linewidth]{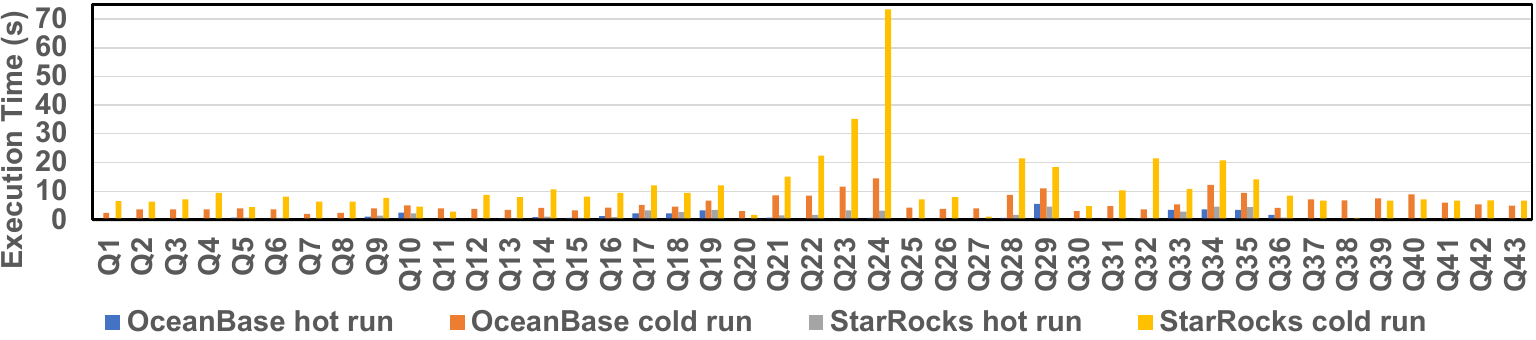}
  \caption{Query performance of OceanBase and StarRocks with ClickBench}
  \label{fig:OBStarRocksClickBench}
  \vspace{-0pt}
\end{figure*}

Although the introduction emphasizes OLTP, Bacchus's shared storage design also targets HTAP and OLAP workloads; we therefore evaluate \textbf{analytical query performance} using standard OLAP benchmarks. TPC-H~\cite{tpch} and TPC-DS~\cite{tpcds} are widely used for analytical/decision-support workloads, and ClickBench~\cite{clickbench} covers analytical query patterns. We compare Bacchus with StarRocks~\cite{starrocks}, a leading analytical engine, to show that our architecture does not sacrifice OLAP performance. We chose StarRocks because it previously had the best performance in our actual daily application requirements. Each system (OceanBase 4.4.x and StarRocks 3.3.13) uses its \textbf{own} query optimizer and thus its \textbf{own} query plans for the same benchmark SQL; the comparison is end-to-end (same benchmark queries, same data scale, same cold/warm methodology), with each engine using its native optimizer and storage. For OceanBase Bacchus itself, the query optimizer and plan generation are unchanged from OceanBase; relative to a non-shared-storage OceanBase, \textbf{storage is the main change} (shared storage, object storage, multi-layer caching), so the TPC-H and TPC-DS results reflect the impact of the new storage layer under the same query plans within OceanBase. Table~\ref{tab:OBvsStarRocks} presents the execution time of 22 complex queries in the TPC-H 100GB benchmark. 
Figure~\ref{fig:OBStarRocksTPCH} shows the results for TPC-H 100GB and 1TB. Figure~\ref{fig:OBStarRocksTPCDS100GB} and Figure~\ref{fig:OBStarRocksClickBench} illustrate the results for TPC-DS 100GB (99 queries) and ClickBench (43 queries). We use both cold and hot runs (hot: data preloaded into cache; cold: no preloading) and compare total query time (hot + cold).

As shown in Figure~\ref{fig:OBStarRocksTPCH}(a), for TPC-H 100GB, OceanBase Bacchus achieved approximately 48.5\% faster query execution than StarRocks, with about one-third of queries nearly doubling in speed (>90\% acceleration). The most significant improvement was observed in query Q14 (327.93\%), while Q9 showed the worst performance with a 35.71\% slowdown. Q9 assesses the aggregated profit of ordered parts per country and per year and combines grouping, sorting, aggregation, and subquery operations, whose complexity may contribute to performance degradation in OceanBase. As illustrated in Figure~\ref{fig:OBStarRocksTPCH}(b), for TPC-H 1TB, performance improved by approximately 40\% in cold runs, while in hot runs there was only a marginal improvement (worst also in Q9, speedup 3.0\%). For TPC-DS 100GB, we achieved a 43.8\% execution-time speedup; as shown in Figure~\ref{fig:OBStarRocksTPCDS100GB}, the two databases performed on a par during hot runs, whereas Bacchus clearly outperformed StarRocks during cold runs. Bacchus reached its peak execution time on Q4 (32.67s), which quantifies the incremental sales generated during the promotion period. For ClickBench in Figure~\ref{fig:OBStarRocksClickBench}, it attained an 89\% speedup (the best among all experiments) but was inferior to StarRocks on a small subset of queries.

These results demonstrate that our improvements to shared storage have no negative impact on performance, indicating preliminary success in OceanBase's adaptive optimizations for object storage architecture. OceanBase Bacchus performs comparably to or even better than leading databases in processing comprehensive queries, confirming the feasibility and effectiveness of the system.


\begin{table}[htp]
\caption{Query performance comparison in TPC-H 100GB}
\small
\begin{tabular}{|c|cc|cc|c|}
\hline
\multicolumn{1}{|c|}{\multirow{2}{*}{\textbf{Query}}} & \multicolumn{2}{c|}{\textbf{OceanBase}}                                      & \multicolumn{2}{c|}{\textbf{StarRocks}}                                      & \multicolumn{1}{c|}{\textbf{Improvement}} \\ \cline{2-6} 
\multicolumn{1}{|c|}{}                                & \multicolumn{1}{c|}{\textbf{hot (s)}} & \multicolumn{1}{c|}{\textbf{cold (s)}} & \multicolumn{1}{c|}{\textbf{hot (s)}} & \multicolumn{1}{c|}{\textbf{cold (s)}} & \multicolumn{1}{c|}{\textbf{hot + cold}}  \\ \hline
Q1                                                    & \multicolumn{1}{c|}{2.28}            & 8.83                                  & \multicolumn{1}{c|}{1.66}            & 9.99                                  & 4.86\%                                    \\ \hline
Q2                                                    & \multicolumn{1}{c|}{0.13}            & 5.67                                  & \multicolumn{1}{c|}{0.10}            & 8.59                                  & 49.83\%                                   \\ \hline
Q3                                                    & \multicolumn{1}{c|}{0.68}            & 10.64                                 & \multicolumn{1}{c|}{0.67}            & 15.67                                 & 44.35\%                                   \\ \hline
Q4                                                    & \multicolumn{1}{c|}{0.63}            & 8.86                                  & \multicolumn{1}{c|}{0.45}            & 20.48                                 & 120.55\%                                  \\ \hline
Q5                                                    & \multicolumn{1}{c|}{0.80}            & 16.38                                 & \multicolumn{1}{c|}{0.75}            & 23.05                                 & 38.53\%                                   \\ \hline
Q6                                                    & \multicolumn{1}{c|}{0.03}            & 5.21                                  & \multicolumn{1}{c|}{0.05}            & 7.54                                  & 44.85\%                                   \\ \hline
Q7                                                    & \multicolumn{1}{c|}{0.82}            & 13.44                                 & \multicolumn{1}{c|}{0.77}            & 18.88                                 & 37.80\%                                   \\ \hline
Q8                                                    & \multicolumn{1}{c|}{0.78}            & 16.91                                 & \multicolumn{1}{c|}{0.57}            & 31.29                                 & 80.10\%                                   \\ \hline
Q9                                                    & \multicolumn{1}{c|}{2.16}            & 21.64                                 & \multicolumn{1}{c|}{1.77}            & 13.53                                 & -35.71\%                                  \\ \hline
Q10                                                   & \multicolumn{1}{c|}{0.76}            & 10.46                                 & \multicolumn{1}{c|}{0.99}            & 24.03                                 & 122.99\%                                  \\ \hline
Q11                                                   & \multicolumn{1}{c|}{0.29}            & 4.36                                  & \multicolumn{1}{c|}{0.20}            & 7.63                                  & 68.39\%                                   \\ \hline
Q12                                                   & \multicolumn{1}{c|}{0.37}            & 10.63                                 & \multicolumn{1}{c|}{0.19}            & 17.10                                 & 57.18\%                                   \\ \hline
Q13                                                   & \multicolumn{1}{c|}{1.23}            & 7.21                                  & \multicolumn{1}{c|}{1.30}            & 9.97                                  & 33.53\%                                   \\ \hline
Q14                                                   & \multicolumn{1}{c|}{0.21}            & 1.58                                  & \multicolumn{1}{c|}{0.16}            & 7.50                                  & 327.93\%                                  \\ \hline
Q15                                                   & \multicolumn{1}{c|}{0.16}            & 2.21                                  & \multicolumn{1}{c|}{0.11}            & 7.60                                  & 225.32\%                                  \\ \hline
Q16                                                   & \multicolumn{1}{c|}{0.45}            & 3.62                                  & \multicolumn{1}{c|}{0.20}            & 7.70                                  & 94.10\%                                   \\ \hline
Q17                                                   & \multicolumn{1}{c|}{0.31}            & 9.37                                  & \multicolumn{1}{c|}{0.32}            & 10.20                                 & 8.68\%                                    \\ \hline
Q18                                                   & \multicolumn{1}{c|}{4.38}            & 11.06                                 & \multicolumn{1}{c|}{2.07}            & 21.16                                 & 50.45\%                                   \\ \hline
Q19                                                   & \multicolumn{1}{c|}{0.44}            & 9.94                                  & \multicolumn{1}{c|}{0.26}            & 10.54                                 & 4.05\%                                    \\ \hline
Q20                                                   & \multicolumn{1}{c|}{0.24}            & 8.89                                  & \multicolumn{1}{c|}{0.17}            & 9.15                                  & 2.08\%                                    \\ \hline
Q21                                                   & \multicolumn{1}{c|}{1.80}            & 13.14                                 & \multicolumn{1}{c|}{1.33}            & 27.54                                 & 93.24\%                                   \\ \hline
Q22                                                   & \multicolumn{1}{c|}{0.29}            & 3.93                                  & \multicolumn{1}{c|}{0.24}            & 8.17                                  & 99.29\%                                   \\ \hline
Total                                           & \multicolumn{1}{c|}{19.24}           & 203.98                                & \multicolumn{1}{c|}{14.33}           & 317.31                                & 48.57\%                                   \\ \hline
\end{tabular}
\label{tab:OBvsStarRocks}
\end{table}

\subsection{Multi-Layer Cache Hit Ratio in Production} \label{sec:cachehit}
To validate the effectiveness of Bacchus's multi-layer caching design (Section~\ref{sec:caching}) under real-world workloads, we collected cache hit ratio metrics from two production deployments over a one-week observation period. We report three metrics: 1) overall cache hit ratio, 2) shared macro-block hit ratio, and 3) private macro-block hit ratio. The first deployment runs an \textbf{OLTP (transactional)} workload; the second runs an \textbf{HTAP (hybrid transactional and analytical)} workload that mixes OLTP and OLAP queries. For OLAP-oriented queries, latency requirements are less strict than for OLTP for two reasons: (1) OLAP typically performs sequential, large-block reads that demand high bandwidth, and object storage is well-suited to large sequential reads and offers high bandwidth; (2) OLAP's response time requirements are relatively less strict than OLTP's. To trade off performance and cost, the system allows some reads to be served from object storage rather than caching all data, so a portion of object storage access is expected in the HTAP case.

\begin{figure}[htp]
  \centering
  \includegraphics[width=0.95\linewidth]{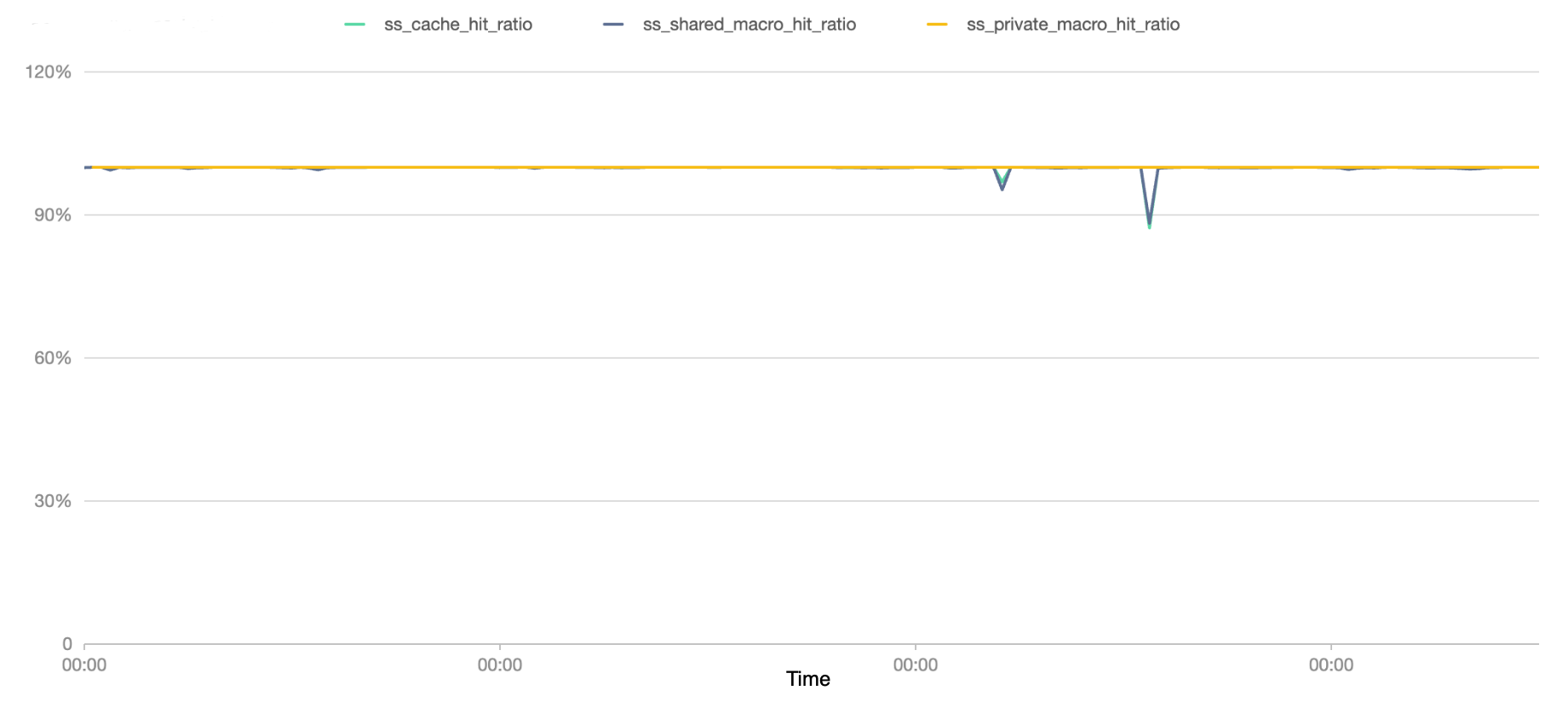}
  \caption{Cache hit ratios for the first production workload (OLTP) over the observation period.}
  \Description{Private macro-block cache holds 100\%; shared macro and overall ratios stay near 100\% with only brief dips; OLTP workload.}
  \label{fig:DBCacheHit}
  \vspace{-0pt}
\end{figure}

Figure~\ref{fig:DBCacheHit} shows the cache hit ratios for the \textbf{first production workload (OLTP)} over the observation period. The private macro-block cache maintains a 100\% hit ratio throughout, indicating that data hot on individual compute nodes is effectively retained in local persistent cache. The shared macro-block cache and the overall cache hit ratio remain near 100\%, with only brief, shallow dips during workload shifts. The upper layers absorb micro-cache misses, so the aggregate cache hit ratio stays high and access to object storage is minimized. The trace demonstrates that the multi-layer design sustains high performance for this OLTP workload across varying conditions.

\begin{figure}[htp]
  \centering
  \includegraphics[width=0.95\linewidth]{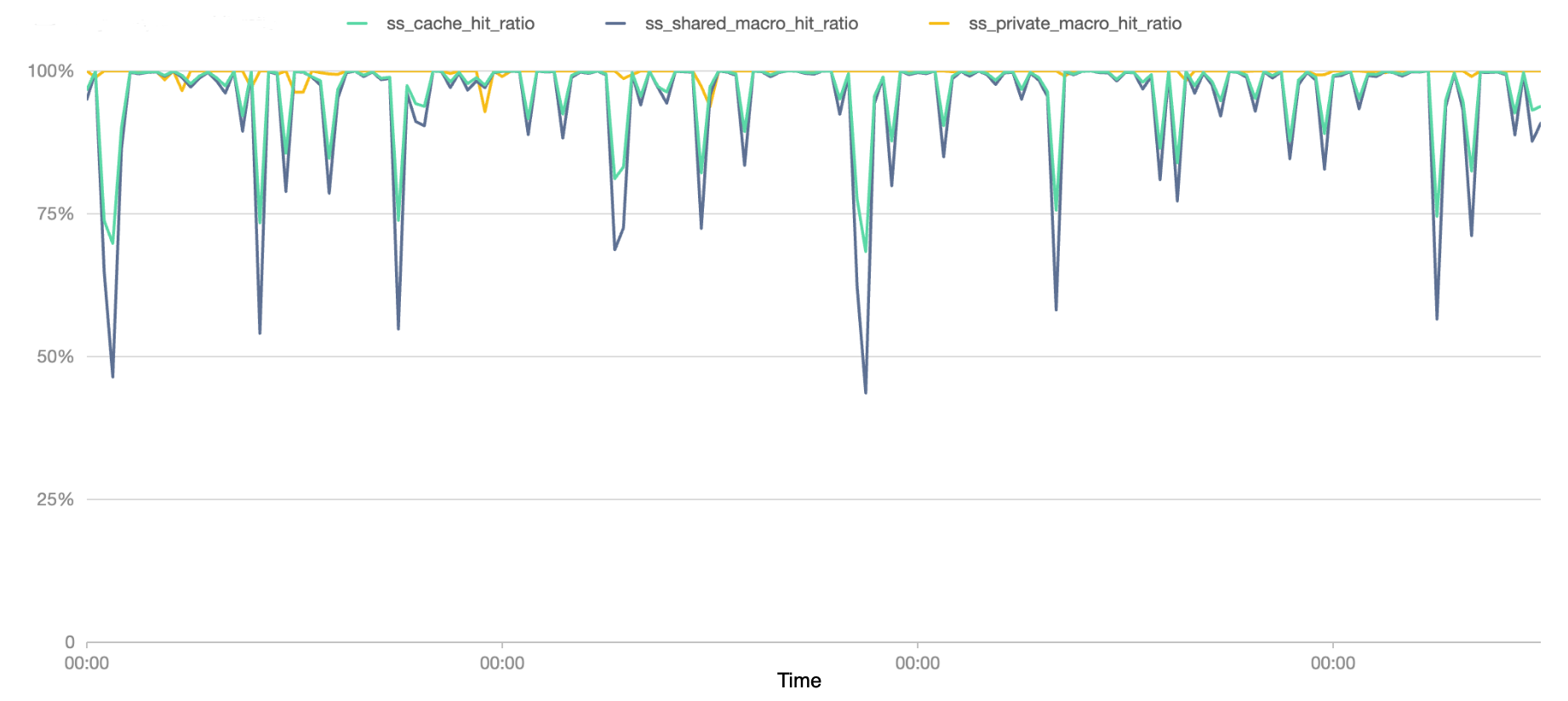}
  \caption{Cache hit ratios for the second production workload (HTAP) over the observation period.}
  \Description{Overall and macro-level hit ratios remain high; private macro cache is the most stable; some dips correspond to OLAP or cold reads from object storage in this HTAP workload.}
  \label{fig:CBCacheHit}
  \vspace{-0pt}
\end{figure}

Figure~\ref{fig:CBCacheHit} presents the same metrics for the \textbf{second production workload (HTAP)} over the same period. This deployment runs a mix of OLTP and OLAP queries, similar to HTAP-style applications. The overall and macro-level hit ratios remain high for most of the time. The private macro cache is the most stable, often at 100\%, while the shared macro and overall ratios track each other with occasional dips that recover quickly. Some of these dips correspond to OLAP queries or cold reads: for OLAP-oriented queries, latency requirements are less strict, and the system intentionally trades off performance and cost by serving a portion of reads from object storage rather than caching all data. The results show that Bacchus's three-tier caching and cache warming (Section~\ref{sec:caching}) effectively serve production HTAP workloads---hot and OLTP-critical data are served from cache with high hit ratios, while the system tolerates some object storage access for OLAP queries to balance performance and cost.

\subsection{OB Shared-storage vs. Shared-nothing} \label{sec:obmulticloud}

\begin{figure}[htp]
  \centering
  \includegraphics[width=0.95\linewidth]{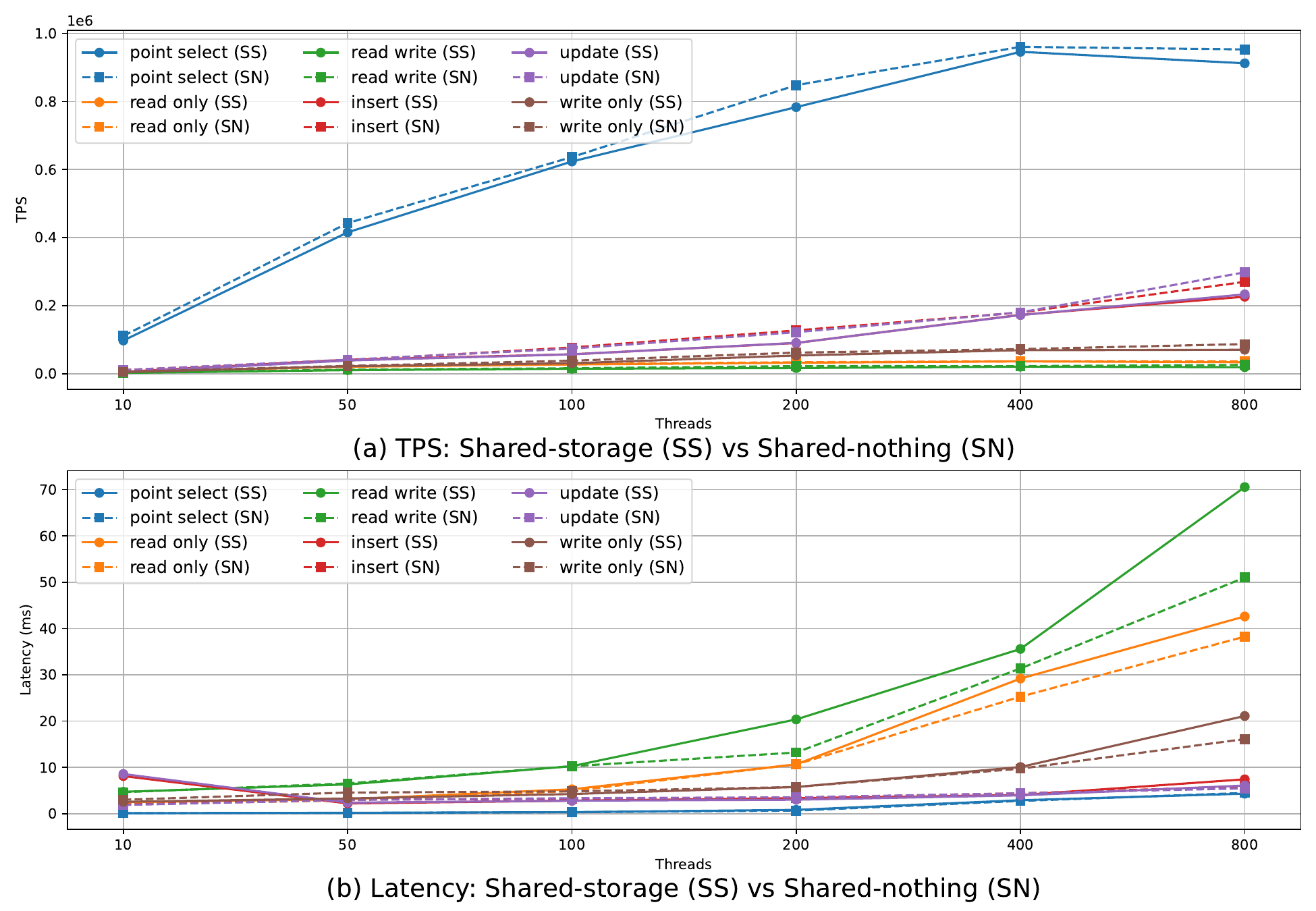}
  \caption{OceanBase in Alibaba Cloud}
  \label{fig:OBSysBench}
  \vspace{-0pt}
\end{figure}

In Alibaba Cloud, we compared the performance of OceanBase under two architectures: Shared-storage (SS) and Shared-nothing (SN).
The dataset was generated using SysBench and consisted of 30 tables, each containing 500,000 rows. We evaluated performance under 6 different workload scenarios, including read–write, insert, and update operations.
As illustrated in Figure~\ref{fig:OBSysBench}, the throughput (TPS) of SS and SN remains comparable across different thread counts. This indicates that although invoking external shared-storage services may introduce additional latency, Bacchus has effectively minimized this impact.

\subsection{Storage Cost Comparison}
We compared storage costs between Bacchus and previous versions of OceanBase (Paetica). Before Bacchus, Paetica adopted a tightly integrated storage-compute architecture, and we typically chose EBS (GP2), a general-purpose storage type provided by Amazon Web Services (AWS), as the storage medium. Now, Bacchus uses a disaggregated compute--storage architecture, using S3 for data persistence while keeping EBS dedicated to cache disks. As shown in Table~\ref{tab:OBStorageCosts}, we compared the two architectures with a storage size of 100 TB.

\textbf{Why Bacchus reduces the number of replicas.} In shared-nothing architectures, the database maintains multiple full replicas of the data (e.g., three) across compute nodes for fault tolerance and distributed consensus; storage cost therefore scales with replica count. In Bacchus's shared-storage architecture, data resides in a \emph{single} shared store (object storage). Durability and availability are provided by the storage layer (object storage is typically replicated by the cloud provider) and by the PALF log service, which has its own replication for the write-ahead log. Compute nodes are stateless and read from the shared store, so the database does not need to maintain multiple full data replicas across nodes---one copy of the data in shared storage suffices. Bacchus can therefore reduce the replica count to one for the data layer (e.g., in a User Tenant Single Replica Cluster deployment), while the log service retains its own replication for RPO=0 and availability. Thanks to the low cost and high reliability of object storage, Bacchus not only reduces the number of replicas in this way but also lowers storage overhead by 59.0\% in OLTP scenarios and 89.0\% in OLAP scenarios, significantly cutting costs while maintaining high performance. The ``data cache ratio'' in Table~\ref{tab:OBStorageCosts} denotes the ratio of cached data chosen based on business needs (e.g., 10\%).

\begin{table*}[htp]
\centering
\caption{Comparison of storage costs between old and new versions of OceanBase in OLTP and OLAP in 100 TB storage}
\small
\begin{tabular}{|llllc|}
\hline
\multicolumn{5}{|c|}{\textbf{OLTP}}                                                                                                                                                                                                                                               \\ \hline
\multicolumn{1}{|l|}{\textbf{Storage Architecture}}                        & \multicolumn{1}{l|}{\textbf{Storage Type}}                         & \multicolumn{1}{l|}{\textbf{Monthly Cost per GB}} & \multicolumn{1}{l|}{\textbf{Replica Count}} & \textbf{Monthly Total Cost} \\ \hline
\multicolumn{1}{|l|}{\textbf{OceanBase (Shared-nothing)}}                  & \multicolumn{1}{l|}{EBS (GP2) Cloud Disk}                          & \multicolumn{1}{c|}{\$ 0.10}                      & \multicolumn{1}{c|}{x3}                     & \$ 30,000                   \\ \hline
\multicolumn{1}{|c|}{\multirow{2}{*}{\textbf{OceanBase (Shared-storage)}}} & \multicolumn{1}{l|}{EBS (GP2)}                                     & \multicolumn{1}{c|}{\$ 0.10}                      & \multicolumn{1}{c|}{\multirow{2}{*}{x1}}    & \$ 10,000                   \\ \cline{2-3} \cline{5-5} 
\multicolumn{1}{|c|}{}                                                     & \multicolumn{1}{l|}{S3 Standard}                                   & \multicolumn{1}{c|}{\$ 0.023}                     & \multicolumn{1}{c|}{}                       & \$ 2,300                    \\ \hline
\multicolumn{1}{|l|}{\textbf{Storage Cost Savings}}                        & \multicolumn{1}{l|}{}                                              & \multicolumn{1}{l|}{}                             & \multicolumn{1}{c|}{}                       & \textit{\textbf{59.0\%}}    \\ \hline
\multicolumn{5}{|c|}{\textbf{OLAP}}                                                                                                                                                                                                                                               \\ \hline
\multicolumn{1}{|l|}{\textbf{Storage Architecture}}                        & \multicolumn{1}{l|}{\textbf{Storage Type}}                         & \multicolumn{1}{l|}{\textbf{Monthly Cost per GB}} & \multicolumn{1}{l|}{\textbf{Replica Count}} & \textbf{Monthly Total Cost} \\ \hline
\multicolumn{1}{|l|}{\textbf{OceanBase (Shared-nothing)}}                  & \multicolumn{1}{l|}{EBS (GP2) Cloud Disk}                          & \multicolumn{1}{c|}{\$ 0.10}                      & \multicolumn{1}{c|}{x3}                     & \$ 30,000                   \\ \hline
\multicolumn{1}{|l|}{\multirow{2}{*}{\textbf{OceanBase (Shared-storage)}}} & \multicolumn{1}{l|}{EBS (GP2) (e.g., data cache ratio 10\%)} & \multicolumn{1}{c|}{\$ 0.10}                      & \multicolumn{1}{c|}{\multirow{2}{*}{x1}}    & \$ 1,000                    \\ \cline{2-3} \cline{5-5} 
\multicolumn{1}{|l|}{}                                                     & \multicolumn{1}{l|}{S3 Standard (pay-as-you-go)}                   & \multicolumn{1}{c|}{\$ 0.023}                     & \multicolumn{1}{c|}{}                       & \$ 2,300                    \\ \hline
\multicolumn{1}{|l|}{\textbf{Storage Cost Savings}}                        & \multicolumn{1}{c|}{}                                              & \multicolumn{1}{c|}{}                             & \multicolumn{1}{c|}{}                       & \textit{\textbf{89.0\%}}    \\ \hline
\end{tabular}
\label{tab:OBStorageCosts}
\end{table*}

\section{Lessons in Practice} \label{sec:disc}
We summarize the lessons learned from the design and practice of shared storage in OceanBase Bacchus.

\textbf{Lesson 1}. Object storage has limited IOPS (Input/Output Operations Per Second) and bandwidth, along with high latency. It is crucial to avoid frequent small read/write operations and to aggregate small objects before storage, rather than storing them individually.

\textbf{Lesson 2}. We configure a separate bucket for each cluster or tenant to leverage the I/O isolation and billing capabilities of object storage buckets.

\textbf{Lesson 3}. Multiple separation strategies are critical: storage and computation separation, read and write separation, log and data separation, hierarchical caching and data separation, and front-end and back-end separation. Service-oriented and pooled resources are equally important.

\textbf{Lesson 4}. The log service is crucial, which ensures statelessness of compute nodes and high performance of OLTP, with RPO=0.

\textbf{Lesson 5}. Combining local micro-block caching with shared macro-block caching is effective. Local micro-block caching optimizes OLTP point queries, while shared macro-block caching enables statelessness of compute nodes and enhances node scaling capabilities.

\textbf{Lesson 6}. Various methods are employed to improve cache hit rates, including cache warming, automatic identification of hot data, elastic scaling of cache, and user-defined rules for hot data.

\textbf{Lesson 7}. Hierarchical centralized management of metadata based on log services achieves metadata aggregation and provides efficient transaction processing and query capabilities. This approach significantly simplifies system implementation, especially for snapshot-related functions, and enhances metadata access performance.

\textbf{Lesson 8}. The unified multi-cloud native shared storage architecture offers cost-effective capabilities for OLTP, OLAP, and KV workloads.

\section{Related Work} \label{sec:work}

\subsection{Shared-storage Databases}

Aurora~\cite{10.1145/3035918.3056101} is a quorum-based~\cite{10.1145/800215.806583} cloud-native relational database. It employs an innovative ``the log is the database'' architecture that decouples log persistence from page generation by transmitting redo logs to the storage layer. Storage nodes asynchronously replay these logs to generate pages, reducing network overhead between compute and storage nodes without relying on specialized hardware. Compute nodes push log streams through a dedicated network to a distributed, durable, and self-healing shared storage volume that spans multiple availability zones. The storage layer asynchronously applies log records to data pages and handles background tasks such as garbage collection, data backup, and recovery, thereby significantly reducing the burden on compute nodes.

Within the canonical cloud-native architecture of ``single-writer, multi-reader + shared storage'', PolarDB~\cite{10.14778/3611540.3611562} leverages a hierarchical modification tracker, linear Lamport timestamps, and an RDMA-based network to allow read-only nodes to obtain globally strong-consistent reads at negligible overhead without waiting for log replay. Built on PolarFS~\cite{10.14778/3229863.3229872}, a distributed file system with a POSIX-like interface for shared storage cloud databases, PolarDB achieves ultra-low latency and high availability. PolarDB later evolved into a multi-primary cloud-native database called PolarDB-MP~\cite{10.1145/3626246.3653377}, which leverages both disaggregated shared memory and storage. PolarDB-MP enables multiple read-write nodes to process transactions concurrently while preserving outstanding performance in high-conflict scenarios.

Databricks Lakehouse~\cite{databricks} is built on open direct-access data formats (e.g., Apache Parquet), persists data in low-cost object storage, and overlays it with a transactional metadata layer that implements ACID transactions, versioning, and other management features. It addresses issues such as staleness, reliability, total cost of ownership, and data lock-in in conventional data warehouses.

Unlike Aurora and PolarDB, which use B+-tree structures and are tightly coupled to specific cloud platforms, Bacchus employs an LSM-tree architecture optimized for object storage, enabling multi-cloud deployment~\cite{xu2025oceanbase} and achieving RPO=0 through service-oriented shared logging. Problems, challenges, and opportunities in native distributed databases are surveyed in a recent tutorial~\cite{xu2024native}. Compared to Databricks Lakehouse, which focuses on analytical workloads, Bacchus targets OLTP workloads with distinct hot/cold data patterns, leveraging multi-layer caching to bridge the performance gap between object storage and OLTP requirements.

\subsection{LSM-tree--based vs. B+-tree--based Databases}
The LSM-tree~\cite{10.1007/s002360050048} employs a hierarchical storage structure, with core components including the in-memory MemTable and the on-disk SSTables. During writes, data is first written sequentially to MemTable, with a backup in the WAL. Once MemTable is full, data is flushed to disk as an immutable SSTable file using sequential I/O. The LSM-tree leverages in-memory buffering and sequential writes to disk to avoid random I/O, significantly enhancing write performance. Common databases based on LSM-tree include LevelDB~\cite{leveldb}, RocksDB~\cite{10.1145/3483840}, Cassandra~\cite{cassandra}, and TiDB~\cite{10.14778/3415478.3415535}.

The B+ tree~\cite{10.1145/356770.356776} is a balanced multiway search tree, with data stored in fixed-size pages (typically 4KB) forming a hierarchical index. Updates are made in place on the original data pages, requiring maintenance of tree balance but potentially triggering page splits or merges. Efficient reads involve binary search from the root node to the leaf nodes, with a stable query complexity of $O(\log N)$. Moreover, leaf nodes are linked sequentially via pointers, enabling efficient range scans~\cite{silberschatz2011database}. Popular B+-tree-based databases include MySQL (InnoDB)~\cite{mysql}, WiredTiger~\cite{wiredtiger}, and PostgreSQL~\cite{postgresql}.

The LSM-tree excels in write performance, far surpassing B+ trees and making it ideal for write-intensive scenarios. It naturally supports high throughput and large-scale data expansion. However, its disadvantages include read amplification, where reads may require querying multiple levels of files. Write amplification is also a concern, with actual disk writes significantly exceeding the volume of user data during compaction. Additionally, poorly designed compaction strategies consume significant CPU, I/O, and disk bandwidth, leading to performance bottlenecks and wasted space, requiring careful tuning to match workload~\cite{10.1007/s00778-019-00555-y}. In contrast, B+ trees offer excellent and predictable read performance, making them suitable for read-heavy scenarios. They are widely adopted in traditional relational databases. However, B+ trees suffer from random I/O and page splits, leading to performance degradation under high-concurrency writes. Furthermore, nodes are often not fully utilized (e.g., maintaining 50\% free space), and compression requires page alignment (e.g., 4KB blocks), which can result in storage fragmentation with frequent updates. Recent studies have extensively focused on narrowing the performance gap between the B+-tree and the LSM-tree~\cite{DBLP:conf/fast/QiaoCZLL022, 10.1145/3633475}.

While existing LSM-tree databases (LevelDB, RocksDB, Cassandra, TiDB) are designed for local or shared-nothing architectures, Bacchus uniquely combines LSM-tree with object storage in a shared storage architecture, solving the cross-node shared-log coordination problem through service-oriented PALF-based logging and achieving stateless compute nodes through multi-layer caching and asynchronous background services.

\textbf{Comparison with Distributed LSM-tree Databases.} TiDB~\cite{10.14778/3415478.3415535} and CockroachDB~\cite{10.1145/3318464.3386134} are distributed LSM-tree databases that employ shared-nothing architectures with data replication across nodes. Both systems maintain metadata (e.g., table schemas, partition information, SSTable manifests) locally on each node, requiring complex distributed consensus protocols (Raft) to synchronize metadata changes across replicas. In contrast, Bacchus introduces a hierarchical metadata management system with SSLog (Shared Storage Log) that centralizes metadata updates in a service-oriented log, eliminating the need for per-node metadata replication. This design enables RO nodes to poll SSLog for metadata updates without participating in consensus protocols, significantly reducing metadata synchronization overhead. Furthermore, Bacchus's PALF-based logging provides stronger consistency guarantees than traditional Raft-based replication: PALF leverages Paxos with optimized batching and pipelining, achieving lower latency for log persistence while maintaining durability across availability zones. Unlike TiDB and CockroachDB, where metadata changes require multi-round Raft consensus, Bacchus's SSLog aggregates metadata updates and flushes them asynchronously, reducing I/O overhead for frequent metadata operations (e.g., compaction metadata, schema changes). This metadata management innovation is particularly beneficial in shared storage architectures where compute nodes are stateless and can be dynamically scaled, as metadata consistency is maintained through the shared log rather than distributed state machines.

\subsection{Shared Log Stream}

The core idea of LFS (Log-Structured File System)~\cite{10.1145/146941.146943} is to treat the entire disk as a circular log stream, where all write operations, including both data and metadata, are appended to the end of the log. Although initially designed for single-machine use, its fundamental concept of a log stream has profoundly influenced subsequent designs of distributed file systems and shared storage.

Corfu~\cite{10.1145/2535930, 10.5555/2228298.2228300} explicitly proposes a distributed shared log based on shared storage for flash clusters. It decouples the storage and management of logs from compute nodes, allowing multiple clients (potentially on different machines) to concurrently append records to this global shared log. Corfu achieves high performance through a mechanism called \textit{Projection}. Tango~\cite{10.1145/2517349.2522732} builds a distributed programming framework backed by the Corfu shared log. It demonstrates how to use the shared log to provide key properties of data management and how to make metadata services more highly available and strongly consistent.

SingleStore~\cite{10.1145/3514221.3526055} has a storage-compute disaggregated architecture, with its logs stored in blob storage~\cite{blob}. However, it relies on log replication rather than a shared log to maintain consistency for new transactions. Building on Aurora's design, Socrates~\cite{10.1145/3299869.3314047} decouples the log from storage and persists it in an independent shared \textit{XLOG} service. Only the primary compute node appends updates to the log; this single-writer paradigm ensures both low latency and high throughput. All other nodes consume the log asynchronously to keep their data replicas up-to-date.

\noindent\begin{minipage}[t]{\dimexpr\linewidth-4pt\relax}
\sloppy
Unlike Corfu and Tango, which provide general-purpose shared log abstractions, or SingleStore and Socrates, which use B+-tree structures, Bacchus integrates service-oriented shared logging (based on PALF) with an LSM-tree architecture optimized for object storage. This combination enables efficient cross-cluster log sharing, eliminates log redundancy, and achieves RPO=0 while supporting multi-cloud deployment---a unique integration not found in existing systems.
\end{minipage}

\textbf{Multi-Writer Support and Correctness.} A key architectural distinction between Bacchus and Socrates (Azure SQL Hyperscale) lies in their write models. Socrates employs a strict single-writer paradigm where only the primary compute node can append to XLOG, creating a bottleneck for write scalability. In contrast, Bacchus supports multiple concurrent writers through a sharded write model: data is partitioned into log streams, with each log stream having a dedicated leader node responsible for writing CLog entries to PALF. Multiple RW nodes can operate concurrently, each serving as leader for different log streams, enabling horizontal write scaling. This design maintains correctness through several mechanisms: (1) \textit{Partition-level serialization}: writes to the same log stream are serialized through its leader, ensuring ordering within each partition; (2) \textit{Cross-partition transaction coordination}: distributed transactions spanning multiple log streams are coordinated via OceanBase 2PC, where a coordinator manages prepare and commit phases across participating log stream leaders; (3) \textit{PALF consensus}: each log stream's PALF service uses Paxos to ensure strong consistency and durability, with majority quorum guaranteeing that committed logs are durable even if minority replicas fail; (4) \textit{Metadata consistency}: SSLog provides a unified metadata log that all nodes consume, ensuring consistent metadata views across RW and RO nodes. This multi-writer architecture enables Bacchus to achieve higher write throughput than single-writer systems like Socrates, while maintaining ACID guarantees through partition-level ordering and distributed transaction coordination. The correctness of this design is validated by the system's ability to handle concurrent writes across partitions, maintain transaction atomicity across log streams, and ensure consistent metadata propagation through SSLog replay.

\section{Conclusion} \label{sec:conc}
This paper introduces OceanBase Bacchus, a shared storage architecture built on OceanBase that integrates an LSM-tree engine with object storage and augments the system with a suite of shared services to improve performance and reduce cost. Shared service-oriented logging based on PALF persists and replicates the write-ahead log, reducing object storage log-I/O overhead and offloading log synchronization from the database layer. The Shared Block Cache Service (BlockServers) enables cache sharing within an availability zone, increasing cache utilization and fostering stateless compute nodes. Asynchronous background-task services offload heavy workloads to dedicated workers, thereby safeguarding high throughput and high availability in the foreground. Collectively, these designs endow OceanBase Bacchus with low cost, elastic resource scaling, and superior resource utilization. As a cloud-native architecture that balances performance and cost, Bacchus contributes meaningfully to the advancement of public cloud databases.



\bibliographystyle{ACM-Reference-Format}
\bibliography{sample}

\end{document}